\documentclass[twocolumn,secnumarabic,amssymb, nobibnotes, aps, prd]{revtex4-2}

\usepackage{subcaption}
\usepackage{graphicx}
\usepackage{amsmath}
\usepackage{amssymb}
\usepackage{lipsum}
\usepackage{braket}
\usepackage{booktabs}
\usepackage{makecell}
\usepackage{multirow}
\usepackage{ragged2e}

\makeatletter
\long\def\@makecaption#1#2{%
  \par
  \addvspace{10pt}
  \small
  \noindent\justifying
  #1\ #2\par
  \addvspace{5pt}}
\makeatother

\begin{document}

\title{Greybody Factors and Hawking Spectra of Quantum-Corrected Black Holes}%

\author{Moslem Shafiee$^{1}$}%
\email[Contact author: ]{m.shafiee72@sharif.edu}
\author{Ahmad Sheykhi$^{1, 2}$}%
\email[Contact author: ]{asheykhi@shirazu.ac.ir}
\affiliation{$^1$Department of Physics, College of
Science, Shiraz University, Shiraz 71454, Iran\\
$^2$Biruni Observatory, College of Science, Shiraz University, Shiraz
71454, Iran}

\begin{abstract}
We investigate Hawking radiation of a quantum-corrected
Schwarzschild black hole with vacuum polarization and conformal
anomaly effects. The corrected geometry has two horizons, a
non-monotonic Hawking temperature that vanishes at a critical
mass, and a logarithmic entropy correction. We compute greybody
factors, spectra, and emission rates for massless bosonic fields
of various spins and angular momenta. Quantum corrections suppress
low-energy greybody factors and enhance high-energy ones. During
late evaporation, where quantum influences become significant, the
decreasing temperature strongly suppresses emission. In this case,
the higher-spin modes become negligible, and the scalar $s=l=0$
mode dominates. The black hole lifetime is greatly extended as it
asymptotically approaches a stable extremal remnant.

\end{abstract}
\maketitle

\section{Introduction}
Black hole evaporation represents one of the most remarkable
manifestations of the interplay between quantum theory and
classical gravity. The discovery of Hawking radiation showed that
quantum fields propagating in a curved spacetime background can
lead to thermal emission from black holes, implying that black
holes gradually lose mass through quantum processes \cite{i1, i2}.
Although Hawking's framework provides a powerful description of
black hole evaporation, it is expected that quantum corrections
become essential when the black hole approaches the endpoint of
evaporation. In particular, the complete disappearance of the
black hole at the end of evaporation and the nature of the final
state remain among the unresolved issues in black hole physics
\cite{i3, i4}. The final stages of black hole evolution are
expected to be strongly affected by quantum gravitational effects,
where the classical description of spacetime may no longer remain
valid. This has motivated the development of various
quantum-corrected and regular black hole geometries aimed at
addressing fundamental issues such as the resolution of spacetime
singularities and the modification of the black hole evaporation
endpoint. A broad class of modified black hole models has been
investigated in this context, including regular black holes
\cite{i5, i6, i7}, as well as quantum gravity inspired approaches
based on loop quantum gravity \cite{i8, i9, i10}, asymptotically
safe gravity \cite{i11, i12}, generalized uncertainty principles
\cite{i13, i14}, and noncommutative geometry \cite{i15}. These
models often predict deviations from the classical Schwarzschild
scenario, including modified horizon structures, altered
thermodynamic properties, and the possible formation of stable
black hole remnants.

By incorporating quantum vacuum effects associated with vacuum
polarization and conformal anomaly, an effective modification of
the Schwarzschild geometry is obtained \cite{shafiee1}. In this
framework, leading-order corrections to the Einstein field
equations modify the horizon structure of the Schwarzschild black
hole, resulting in the splitting of the classical event horizon
into two distinct horizons: an outer horizon that reduces to the
classical Schwarzschild horizon in the appropriate limit and a
purely quantum inner horizon. Under specific conditions, these
quantum corrections can prevent the formation of a singularity
within the effective description and lead to non-singular black
hole configurations. In such cases, the inner horizon provides a
quantum modification of the internal structure of the black hole,
extending the classical Schwarzschild solution beyond its original
description.

Thermodynamic properties of this quantum-corrected Schwarzschild
black hole have been investigated in detail in our previous study
\cite{shafiee2}. Unlike the classical Schwarzschild solution,
where the Hawking temperature increases monotonically as the black
hole mass decreases and eventually diverges, the quantum-corrected
geometry exhibits a qualitatively different thermodynamic
behavior. The Hawking temperature first increases with decreasing
black hole mass, reaches a maximum value when quantum corrections
become significant near the Planck scale, and subsequently
decreases toward zero as the black hole approaches its minimum
mass. At this final stage, the inner and outer horizons approach
each other and eventually merge, resulting in the formation of an
extremal remnant with vanishing Hawking temperature and a stable
final configuration. Furthermore, the heat capacity changes sign
after the maximum temperature is reached, indicating a
thermodynamic phase transition from an unstable to a stable
configuration. Consequently, the quantum-corrected black hole can
potentially reach thermal equilibrium with its surrounding
environment in the stable regime. In addition, the entropy
acquires a logarithmic correction to the classical
Bekenstein-Hawking area law, reflecting the quantum nature of the
underlying geometry.

The modification of the black hole geometry induced by quantum
effects is expected to influence not only the thermodynamic
properties but also the propagation of quantum fields around the
black hole. In particular, Hawking radiation observed at infinity
is determined by the interplay between the thermal emission near
the outer horizon and the transmission probability of the emitted
particles through the effective potential barrier outside the
horizon. Therefore, any modification of the spacetime structure,
including changes in the horizon configuration and the effective
potential, can leave imprints on the greybody factors, emission
spectra, and total luminosity. This motivates a detailed
investigation of the radiation properties of quantum-corrected
black holes beyond the analysis of their thermodynamic behavior.

The greybody factors provide the essential link between the
Hawking radiation generated near the black hole outer horizon and the
radiation detected by an asymptotic observer. Although the
radiation emitted locally near the outer horizon has a thermal
character, the curvature of spacetime surrounding the black hole
acts as an effective potential barrier that partially reflects the
emitted fields. Consequently, the transmission probability through
this barrier modifies the ideal blackbody spectrum and determines
the actual energy emission rates and luminosity of the black hole.
Since the greybody factors depend on the properties of the
background geometry as well as the spin and angular momentum modes
of the emitted fields, they provide a powerful probe of deviations
from the classical black hole solutions. This dependence has
motivated extensive investigations of black hole radiation in a
wide range of gravitational backgrounds, aiming to understand how
different geometrical and physical properties influence the
observable emission spectrum.

Following the original formulation of Hawking radiation, extensive
studies have been devoted to calculating greybody factors and
emission rates for different black hole backgrounds. In
particular, Page performed detailed numerical calculations of
particle emission rates from Schwarzschild and Kerr black holes
\cite{gb1, gb2, gb3}, establishing the importance of greybody
factors in determining the relative contributions of different
particle species. Subsequent studies extended these analyses to charged
\cite{gb4, gb4b}, rotating \cite{gb5, gb6}, and higher-dimensional
black holes \cite{hb1, hb2, hb3, hb4, hb5, hb6, hb7, hb8, hb9,
hb10}, where the modifications of the effective potential lead to
distinct emission spectra and transmission properties.

Beyond classical solutions, greybody factors have also been
investigated in various modified gravity and quantum gravity
inspired black hole geometries. Studies of regular black holes
\cite{rb1, rb2, rb3, rb4, rb5}, black holes in modified gravity
theories \cite{md1, md2, md3, md4, md5, md6}, and
quantum-corrected backgrounds \cite{qc1} have shown that
corrections to the spacetime geometry can significantly alter the
effective potential and consequently modify the greybody factors
and Hawking spectra. For example, quantum corrections inspired by
the generalized uncertainty principle \cite{qg1, qg2, qg3} and
loop quantum gravity \cite{qg4, qg5, qg6, qg7, qg8} have been
shown to affect the transmission probability of emitted fields,
indicating that greybody factors can serve as signatures of
underlying quantum gravitational effects.

However, despite these developments, the impact of quantum
corrections associated with vacuum polarization and conformal
anomaly on the greybody factors and Hawking emission spectra of
Schwarzschild-like black holes with a modified horizon structure
has not been systematically explored. In particular, the
evaporation process of quantum-corrected black holes approaching
an extremal remnant, where the Hawking temperature decreases and
the radiation becomes strongly suppressed, requires a detailed
analysis of the interplay between the modified geometry, greybody
factors, and particle emission rates.

In this work, we investigate how the quantum corrections encoded
in the modified Schwarzschild geometry affect the greybody
factors, Hawking radiation spectrum, and total emission rates of
massless bosonic fields. By considering scalar, photon, and
graviton perturbations, we analyze the modifications of the
transmission probabilities and the relative contributions of
different spin and angular momentum modes throughout the evolution
of the black hole. In particular, we focus on the late stages of
evaporation, where quantum corrections become significant and the
black hole approaches an extremal remnant. In this regime, our
analysis is performed within an effective semi-classical
framework, in which the quantum-corrected geometry is treated as a
fixed background, allowing us to identify the deviations from the
classical Schwarzschild evaporation behavior induced by the
quantum corrections.

A distinctive feature of the present framework is that the quantum
corrections are derived from quantum vacuum effects, incorporated
through the conformal anomaly and vacuum polarization, rather than
being introduced as independent phenomenological modifications of
the black hole metric. Although various regular black hole models
and quantum gravity inspired scenarios have been proposed to
describe deviations from the classical Schwarzschild solution and
the possible formation of black hole remnants, in many cases the
corresponding corrections are introduced through effective metric
ansatz or model-dependent parameters. In contrast, the present
approach provides an effective semi-classical description in which
the modified geometry is directly related to quantum field
effects.

The organization of this paper is as follows. In Sec. II, we
briefly review the quantum-corrected Schwarzschild geometry and
its thermodynamic properties. In Sec. III, we derive the effective
potentials and calculate the greybody factors for different
massless fields. The Hawking emission spectra and the
contributions of various spin and angular-momentum modes are
studied in Sec. IV. In Sec. V, we analyze the total emitted power
and the evaporation process, including the effect of quantum
corrections on the black hole lifetime. Finally, we summarize our
results and discuss their implications in Sec. VI.

Throughout this work, we adopt natural units such that
$G=c=k_B=\hbar=1$. However, in order to make the scale of quantum
effects explicit, factors of the Planck mass $M_{Pl}$ are retained
in all relevant expressions.
\section{Quantum-corrected Schwarzschild black hole}\label{section 2}
Quantum fluctuations of the vacuum, rooted in the quantum nature
of fields and the Heisenberg uncertainty principle, imply that the
vacuum state can possess a nonzero energy. This provides a strong
motivation for revisiting the solutions of Einstein's equations,
particularly in the presence of quantum fields in the vacuum
state. Within the framework of semi-classical gravity, Einstein's
equations are generalized as
\begin{equation}\label{1}
G_{\mu\nu}=8\pi\braket{T_{\mu\nu}},
\end{equation}
where $\braket {T_{\mu\nu}}$ denotes the expectation value of the
stress-energy tensor of the quantum fields in the vacuum state.
Consequently, one expects the solutions of Einstein's equations,
including the Schwarzschild solution, to be modified by quantum
effects. The expectation value of the quantum stress-energy tensor
depends on the nature of the quantum fields, the quantum state,
and the geometry of spacetime. In general, the relevant quantum
contributions include conformal anomalies, vacuum polarization,
and Hawking radiation \cite{qt1, qt2, qt3}. Taking into account
the effects of conformal anomaly and vacuum polarization, the
Schwarzschild solution has been shown to be modified as
\cite{shafiee1}
\begin{equation}\label{2}
ds^2=-f(r)dt^2+\frac{dr^2}{f(r)}+r^2\left(d\theta^2+\sin^2\theta d\phi^2\right),
\end{equation}
where
\begin{equation}\label{3}
f(r)=1-\frac{2M}{r}+\frac{c M_{Pl}^2}{r^2}
+\frac{c^\prime M_{Pl}^2M}{r^3}
+\frac{c^{\prime\prime}M_{Pl}^2M^2}{r^4}.
\end{equation}
Here, $M$ denotes the black hole mass, while $c$, $c^\prime$, and
$c^{\prime\prime}$ are dimensionless constants whose values depend
on the type and nature of the quantum fields under consideration.
For example, for massless quantum fields with spins $s=0,1,2$,
their values are $c=0.004$, $c^\prime=0.029$, and
$c^{\prime\prime}=1.358$. The first two correction terms arise
from vacuum polarization, whereas the third correction term
results from the combined effects of vacuum polarization and the
conformal anomaly. The function $f(r)$ has two positive real
roots, indicating that quantum corrections modify the black hole
structure such that, instead of a single event horizon, two
horizons are present. The larger root, corresponding to the outer
horizon, is given by
\begin{equation}\label{4}
r_+\simeq\frac{2M}{1+a\left(\dfrac{M_{Pl}}{M}\right)^2},
\quad
a=\frac{c}{4}+\frac{c^\prime}{8}+\frac{c^{\prime\prime}}{16}.
\end{equation}
The outer horizon can thus be regarded as the quantum-corrected
counterpart of the classical Schwarzschild event horizon, with its
radius slightly smaller than the classical value. The inner
horizon, corresponding to the smaller root and having a purely
quantum origin, equals to
\begin{equation}\label{5}
r_-\simeq
\frac{\left(\dfrac{c^{\prime\prime}}{2}\dfrac{M}{M_{Pl}}\right)^{\frac{1}{3}}}
{1-b\left(\dfrac{c^{\prime\prime}}{2}\dfrac{M}{M_{Pl}}\right)^{\frac{1}{3}}\dfrac{M_{Pl}}{M}}L_{Pl},
\quad
b=\frac{2c^\prime+c^{\prime\prime}}{3c^{\prime\prime}}.
\end{equation}
Here, the final expression has been written in terms of the Planck
length. It is worth noting that, for an astrophysical black hole,
the inner horizon can be much larger than the Planck length.

\begin{figure}[t!]
\centering
\includegraphics[scale=0.41]{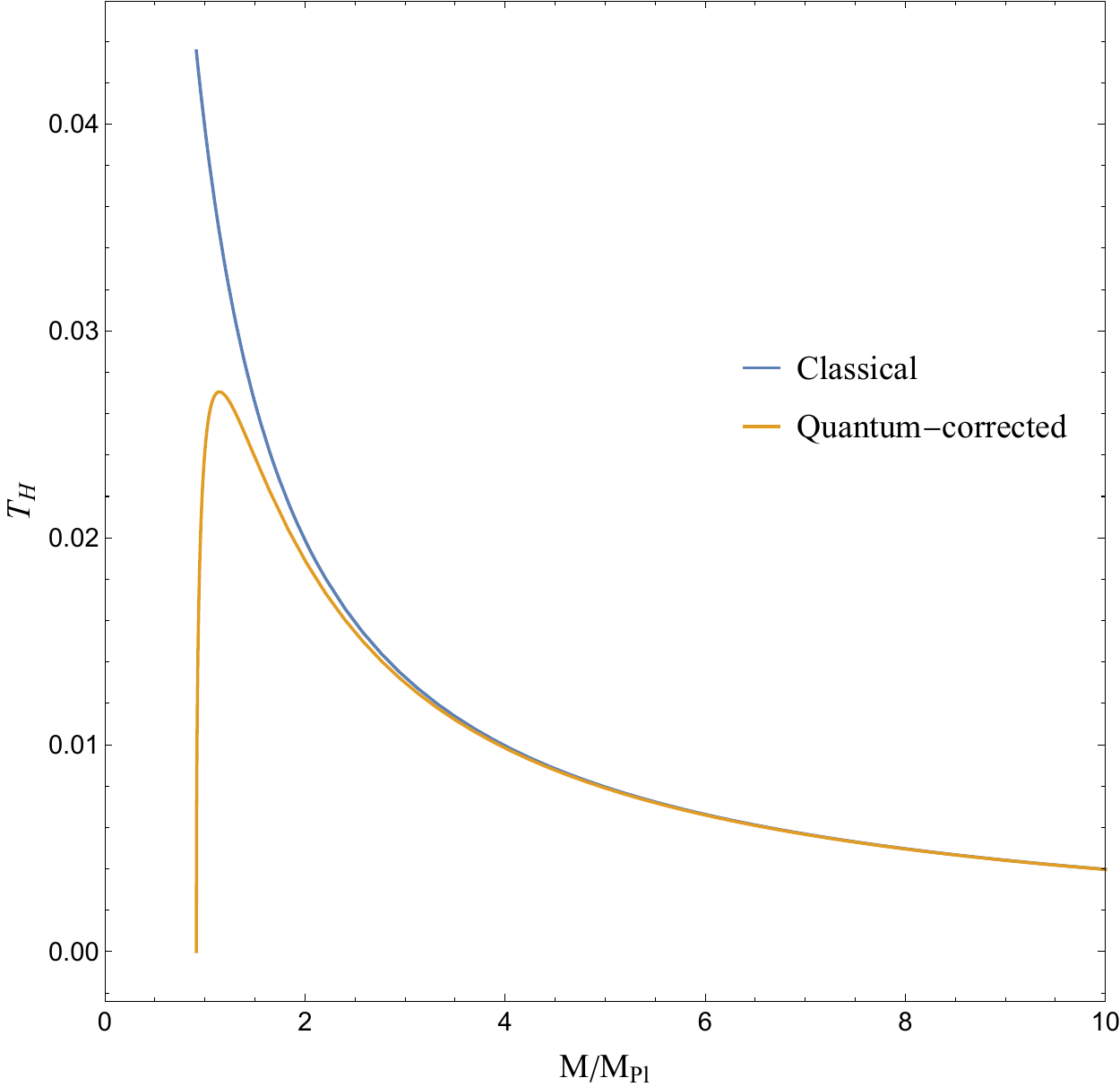}
\caption{This plot compares the Hawking temperature of the black
hole in the classical and semi-classical regimes. While the
classical Schwarzschild temperature increases monotonically as the
black hole mass decreases, the quantum-corrected temperature
reaches a maximum at the critical mass $M_c=1.15 M_{Pl}$. For
smaller masses, the temperature decreases and eventually vanishes
at $M_{\min}=0.92 M_{Pl}$. At this point, the Hawking radiation is
suppressed and the evaporation process terminates, leaving behind
an extremal remnant.}\label{HT}
 \end{figure}

Given the modifications to the black hole structure induced by
quantum corrections, it is natural to expect corresponding changes
in Hawking radiation and the thermodynamic behavior of the black
hole. The Hawking temperature is determined by the surface gravity
at the event horizon \cite{i1, i2}. For the quantum-corrected
black hole described by the metric above, the Hawking temperature
is determined as
\begin{equation}\label{6}
T_H=\frac{1}{4\pi}\left(\frac{df}{dr}\right)\bigg|_{r=r_+}.
\end{equation}
Figure \ref{HT} shows the behavior of the Hawking temperature for
the quantum-corrected black hole and its comparison with the
classical Schwarzschild case. As can be seen, quantum corrections
reduce the Hawking temperature relative to its classical value.
The temperature reaches a maximum with the value of
$T_{max}=38\times10^{29}\mathrm{K}$ at a critical black hole mass,
$M_c=1.15 M_{Pl}$, during the evaporation process. Beyond this
point, the temperature decreases sharply and eventually vanishes
as the black hole mass approaches $M_{\min}=0.92 M_{Pl}$, causing
the Hawking radiation to cease. As the black hole evaporates, the
outer and inner horizons approach each other. At the end of the
evaporation process, the two horizons coincide, forming a single
degenerate horizon, and the trapped region between them disappears
\cite{shafiee2}. Thus, in this picture, the final state of the
black hole evaporation is an extremal remnant with a vanishing
Hawking temperature.
Quantum corrections alter the Hawking temperature, which in turn
modifies other thermodynamic quantities of the black hole,
including its entropy. By applying the first law of black hole
mechanics, the entropy of the quantum-corrected black hole can be
expressed in terms of the area of the outer horizon, i.e., $A=4\pi
r^2_+$. To first order in the quantum corrections, the resulting
entropy is given by \cite{shafiee2}
\begin{equation}\label{7}
S_{BH}=\frac{A}{4}+\frac{\pi}{2}(c^{\prime}+c^{\prime\prime}) \ln \left(\frac{A}{16\pi}\right).
\end{equation}
The first term is the standard Bekenstein-Hawking entropy in the
classical limit, whereas the second term represents a logarithmic
quantum correction. Logarithmic corrections to black hole entropy
have been predicted in a variety of approaches to
quantum-corrected black holes and quantum gravity, including
regular black hole models and loop quantum gravity. This
recurrence across different theoretical frameworks suggests that
logarithmic corrections may represent a generic feature of quantum
gravity effects. A key distinction between the present approach
and several other models is that the logarithmic correction here
arises from the backreaction of quantum fields on the spacetime
geometry, rather than being introduced through phenomenological
assumptions or ad hoc modifications.

Also, an analysis of the black hole heat capacity,
$C=\frac{dM}{dT_{H}}$, predicts a thermodynamically stable state
during the final stages of evaporation. In contrast to the
classical case, in which the black hole becomes hotter as it
radiates ($C<0$), a thermodynamic phase transition occurs when the
black hole temperature reaches its maximum, at which point the
heat capacity changes sign and becomes positive ($C>0$).
Consequently, as the black hole continues to radiate, its
temperature decreases until it eventually vanishes, at which point
Hawking radiation ceases. The black hole therefore exhibits a form
of thermodynamic stability during the final stages of its
evaporation.

In summary, quantum corrections modify both the structure and
thermodynamic behavior of the black hole, leading to a scenario
that differs qualitatively from the classical picture. Rather than
undergoing complete evaporation, the black hole approaches an
extremal remnant with a vanishing temperature and a
thermodynamically stable state at the endpoint of evaporation.

In the next section, we investigate the behavior of massless
bosonic quantum fields in the quantum-corrected Schwarzschild
spacetime and examine how quantum corrections affect the particle
emission rates from the black hole.
\section{Effective potential and greybody factors for quantum-corrected black hole}\label{section 3}
The particle emission spectrum of a black hole is analogous to the
Planck distribution for blackbody radiation. For a black hole with
Hawking temperature $T_H$, the distribution of emitted particles
with spin $s$ can be expressed as \cite{i1, i2}
\begin{equation}\label{8}
n(\omega)=\frac{1}{e^{\omega/T_H}-(-1)^{2s}},
\end{equation}
where $\omega$ denotes the energy of the produced particles.
However, this expression characterizes the emission rate at the
horizon rather than the flux reaching asymptotic infinity. A
fraction of the particles produced near the outer horizon is
backscattered by the spacetime curvature, preventing them from
escaping. Consequently, the physical flux observed at infinity is
modulated by the greybody factor, which encodes the influence of
the black hole geometry and the spin-dependent properties of the
field. Thus, the radiation emitted by a black hole deviates from
an ideal blackbody spectrum, with the greybody factor determining
the energy-dependent attenuation of the emission.
\subsection{Quantum field perturbations and effective potentials}\label{section 3a}
Considering the propagation of a scalar field $\Phi$, its equation of motion in a curved spacetime takes the form
\begin{equation}\label{9}
\left(\Box^{(4)}+m^2\right)\Phi=0,
\end{equation}
where
$\Box^{(4)}=g^{\mu\nu}\bigtriangledown_{\mu}\bigtriangledown_{\nu}$
denotes the four dimensional d'Alembertian, $g_{\mu\nu}$ is the
metric describing the spacetime in Eq. \eqref{2}, and $m$ is the
mass of the scalar field. Using separation of variables, the
solutions of the above equation can be written as \cite{qft}
\begin{equation}\label{10}
\Phi(t, r, \theta, \phi)=\sum_{l=0}^{\infty} \sum_{m=-l}^{l}\int^\infty_0 e^{-i\omega t}\frac{R_ l(r)}{r}\ Y_{lm} (\theta,\phi)d\omega,
\end{equation}
where $Y_{lm} (\theta,\phi)$ are the spherical harmonics and
$\omega$ represents the energy associated with the corresponding
mode. In the above expression, only the outgoing modes are
considered, corresponding to the particles propagating away from
the black hole.

The radial equation, namely the equation satisfied by $R_ l(r)$
can be cast into a Schrodinger-like form by introducing the
tortoise coordinate
\begin{equation}\label{11}
dr^*=\frac{dr}{f(r)}.
\end{equation}
This coordinate maps the outer horizon $r=r_+$ to
$r^*\rightarrow-\infty$, while spatial infinity remains at
$r^*\rightarrow+\infty$, allowing the radial equation to be
interpreted as a one-dimensional scattering problem.

The radial equation then becomes
\begin{equation}\label{12}
\frac{d^2R_ l}{dr^{*2}}+\left(\omega^2-V_l(r)\right)R_ l(r^*)=0.
\end{equation}
Here, $V_l(r)$
denotes the effective potential, which has the form
\begin{equation}\label{13}
V_l(r)=f(r)\left(\frac{f^\prime(r)}{r}+\frac{l(l+1)}{r^2}+m^2\right).
\end{equation}
Consequently, a scalar particle with energy $\omega$ and angular
momentum $l$ emitted by the black hole must overcome this
effective potential in order to reach asymptotic infinity. It
therefore follows that a fraction of the emitted particles can be
captured by the black hole, while only those that penetrate the
effective potential barrier can escape to infinity.

The field equations for massless bosonic particles with higher
spins can, for the cases considered here, be reduced to a scalar
type radial equation with the corresponding spin-dependent
effective potential \cite{price1, price2}. In particular, the
relevant radial modes are obtained by solving the massless
Klein-Gordon equation with the appropriate effective potential,
which can be written in the unified form
\begin{equation}\label{14}
V_{sl}(r)=f(r)\left(\frac{f^\prime(r)}{r}(1-s^2)+\frac{l(l+1)}{r^2}\right),\quad l\geq s.
\end{equation}
For $s=2$, this potential describes the axial gravitational
perturbations (the Regge-Wheeler sector).

Thus, for photons and gravitons, it is sufficient to consider the
corresponding radial equation with their respective spin-dependent
effective potentials for $s=1$ and $s=2$. Introducing the
dimensionless quantities
\begin{equation}\label{15}
x=\frac{r}{M},\quad \Omega=M\omega,\quad \mu=\frac{M_{Pl}}{M},
\end{equation}
the radial equation can be rewritten as
\begin{equation}\label{16}
\frac{d^2R_ {sl}}{dx^{*2}}+\left(\Omega^2-\widetilde{V}_{sl}(x)\right)R_{sl}(x^*)=0,
\end{equation}
where the dimensionless effective potential is defined as
\begin{align}\label{17}
\widetilde{V}_{sl}(x)=M^2V_{sl}(r)=\left(1-\frac{2}{x}+\frac{c\mu^2}{x^2}+\frac{c^\prime\mu^2}{x^3}+\frac{c^{\prime\prime}\mu^2}{x^4}\right)\nonumber\\
\times\left[\left(\frac{2}{x^3}-\frac{2c\mu^2}{x^4}-\frac{3c^\prime\mu^2}{x^5}-\frac{4c^{\prime\prime}\mu^2}{x^6}\right)(1-s^2)+
\frac{l(l+1)}{x^2}\right],
\end{align}
and the dimensionless tortoise coordinate satisfies $x^*=r^*/M$.

The effective potential, as given by Eqs. \eqref{14} and
\eqref{17}, depends on the properties of the emitted particle,
including its angular momentum and spin, as well as on
the geometric properties of the spacetime, such as the black hole
mass and the quantum modifications to the metric.
\begin{figure}[t]
    \centering

    \begin{subfigure}{0.95\columnwidth}
        \centering
        \includegraphics[width=\linewidth]{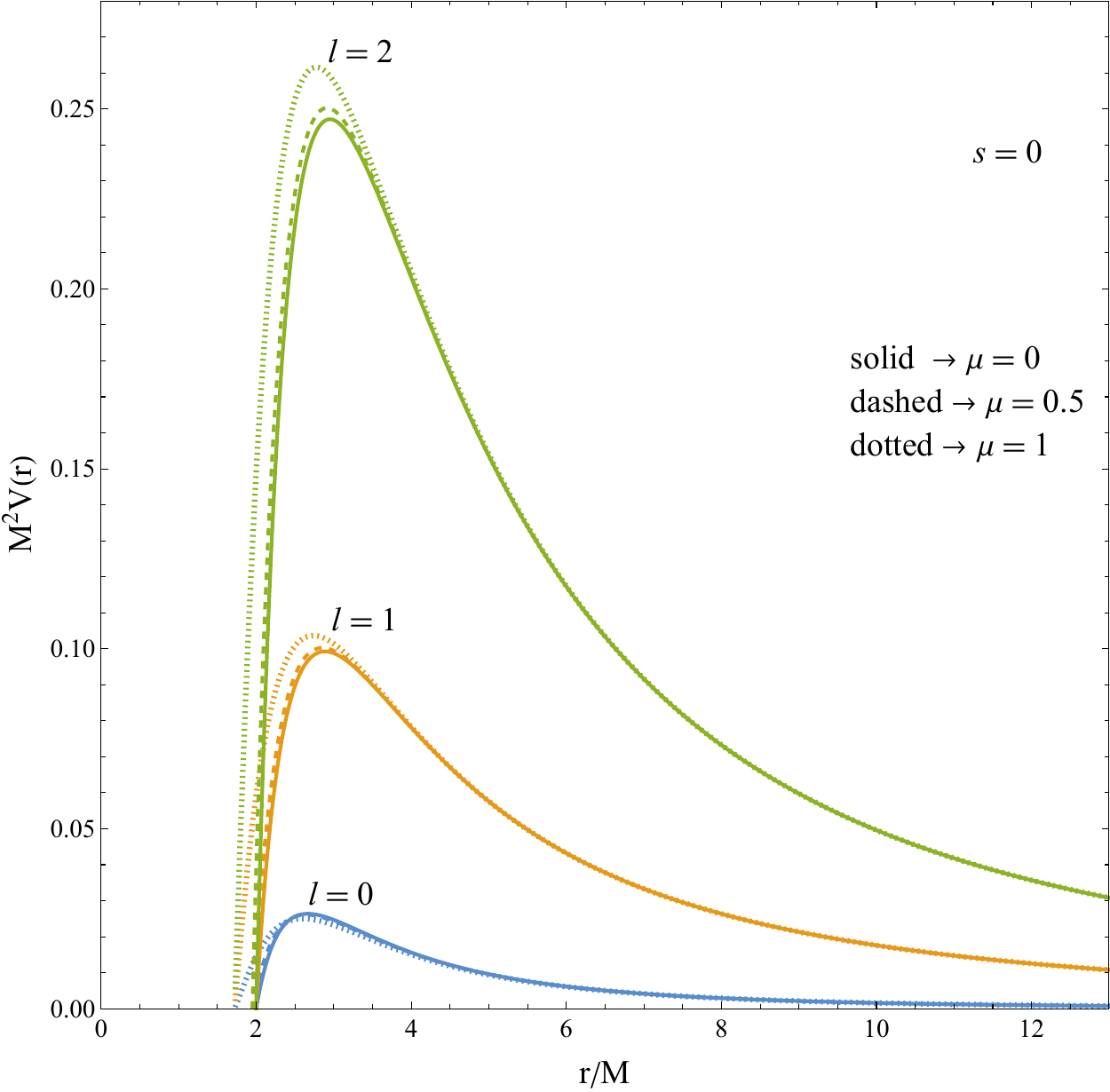}
        \caption{}
        \label{fig:2a}
    \end{subfigure}

    \vspace{0.5em}

    \begin{subfigure}{0.95\columnwidth}
        \centering
        \includegraphics[width=\linewidth]{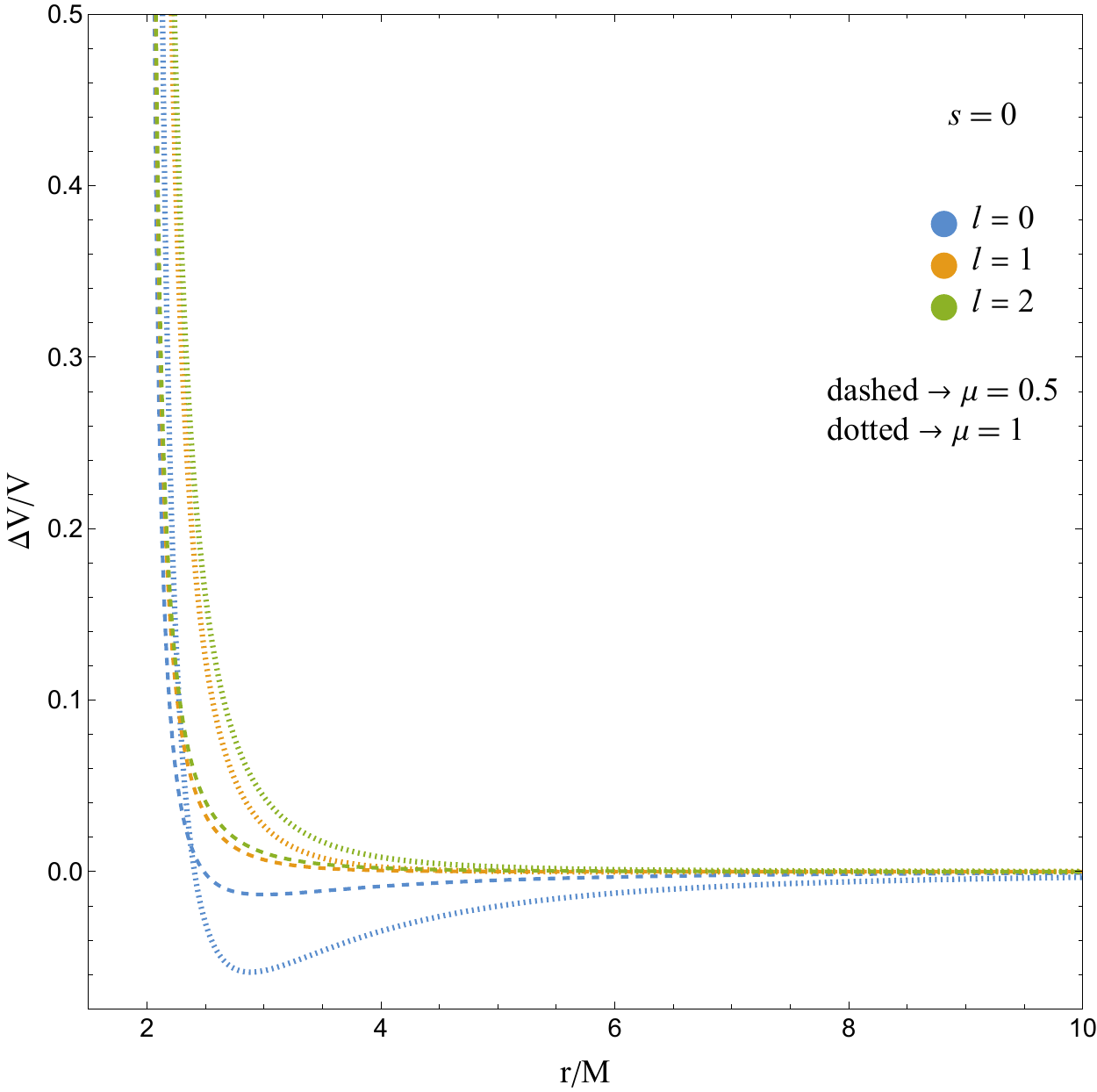}
        \caption{}
        \label{fig:2b}
    \end{subfigure}

\caption{(a) The effective potential for massless scalar particles
for different values of $l$ and $\mu$. (b) Relative difference of
the effective potential between the quantum-corrected cases
$\mu=0.5, 1$ and the classical case $\mu=0$. Quantum corrections
reduce the effective potential for the $s=l=0$ mode, while they
increase it for modes with nonzero angular momentum.}
    \label{fig:2}
\end{figure}

\begin{figure}[t]
    \centering

    \begin{subfigure}{0.95\columnwidth}
        \centering
        \includegraphics[width=\linewidth]{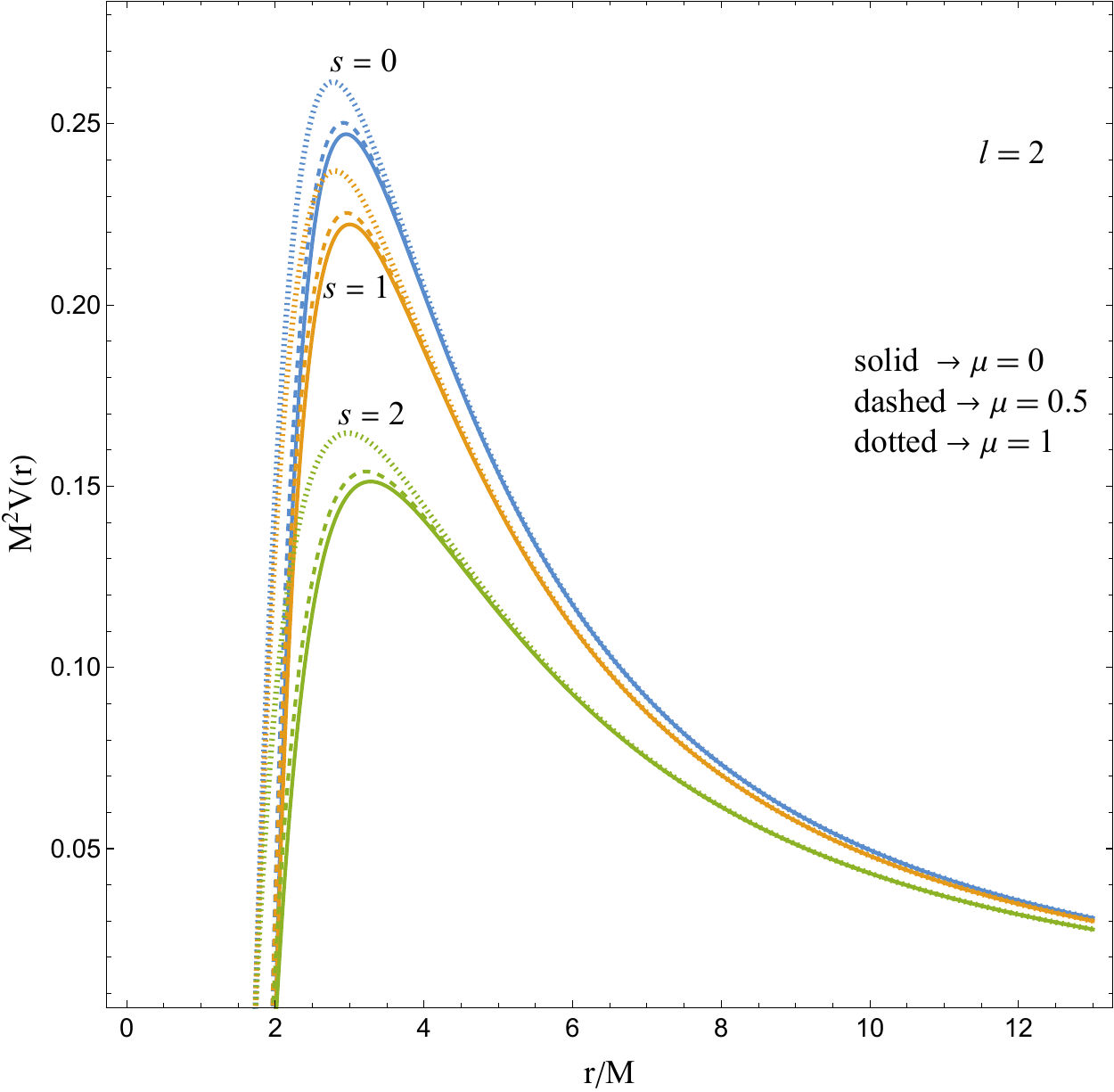}
        \caption{}
        \label{fig:3a}
    \end{subfigure}

    \vspace{0.5em}

    \begin{subfigure}{0.95\columnwidth}
        \centering
        \includegraphics[width=\linewidth]{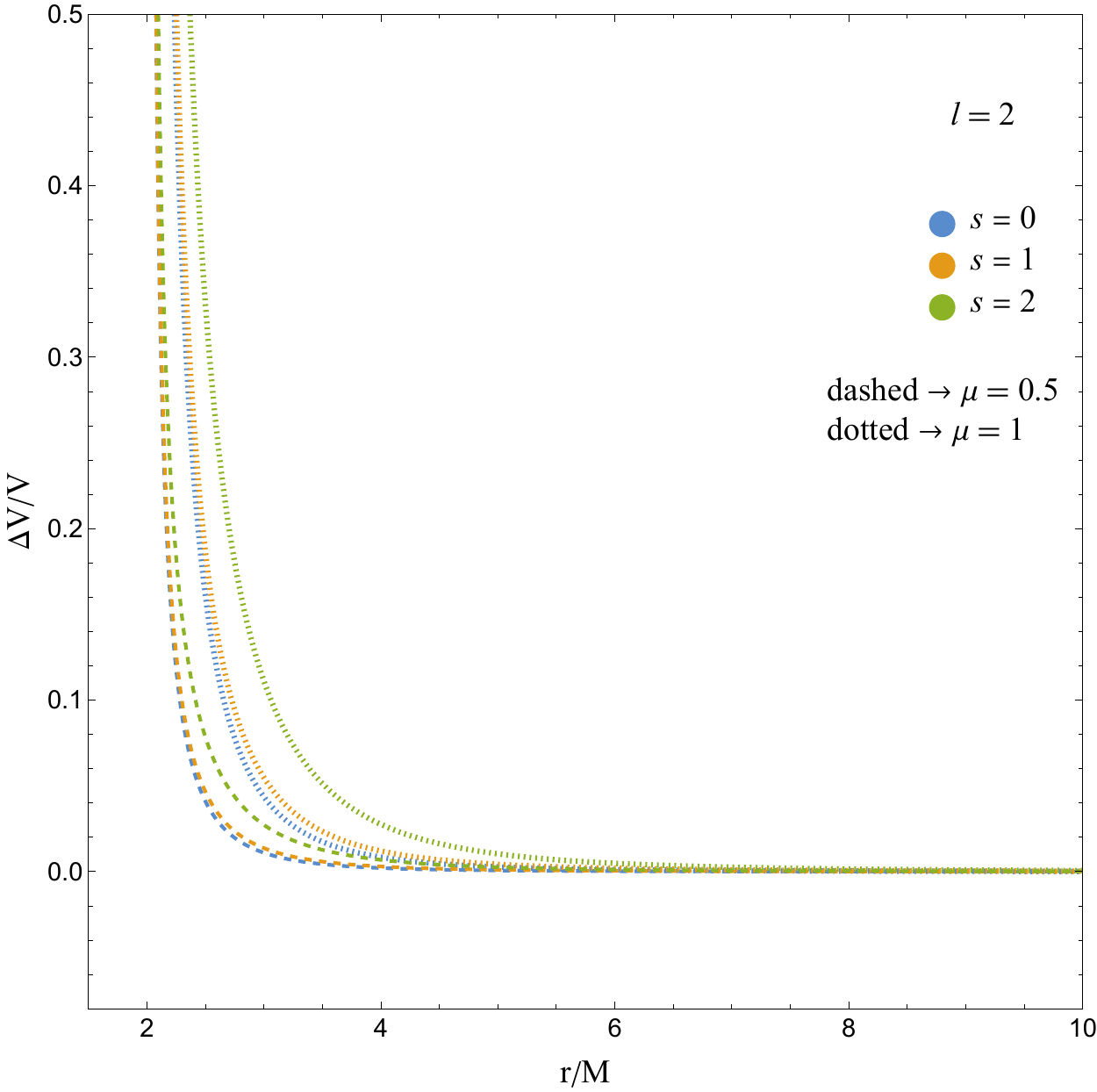}
        \caption{}
        \label{fig:3b}
    \end{subfigure}

\caption{(a) The effective potential for massless bosonic
particles with $l=2$ for different values of $s$ and $\mu$. (b)
Relative difference of the effective potential between the
quantum-corrected cases $\mu=0.5, 1$ and the classical case
$\mu=0$. Quantum corrections increase the effective potential for
the considered nonzero angular momentum modes.}
    \label{fig:3}
\end{figure}
Figure \ref{fig:2a} shows the effective potential for massless
scalars particles for different values of $l$ and $\mu$. As can be
seen, increasing $l$ raises the height of the potential barrier,
indicating that the transmission of modes with larger angular
momentum is increasingly suppressed.

The role of the quantum corrections, however, is somewhat
different. If the black hole is assumed to begin the evaporation
process with a relatively large mass, corresponding to $\mu=0$,
the quantum corrections are negligible and the effective potential
approaches its classical form. As the black hole loses mass and
$\mu$ increases, the quantum corrections become progressively more
significant. For example, at $\mu=0.5$, they already modify the
shape of the effective potential, particularly in the region close
to the black hole. During the late stages of evaporation, when
$\mu=1$, the quantum corrections become most pronounced and
produce the largest deviations from the classical potential.
Figure \ref{fig:2b} shows the relative difference in the effective
potential for $\mu=0.5, 1$ with respect to the classical case,
$\mu=0$. As the black hole mass decreases and the quantum effects
become increasingly significant, the deviations in the effective
potential become more pronounced. In particular, the reduction of
the effective potential for the $s=l=0$ mode and its enhancement
for the other modes become increasingly significant as the quantum
effects grow stronger.

By fixing $l=2$, the influence of the particle spin on the
effective potential can also be examined. As shown in Fig.
\ref{fig:3a}, increasing the spin reduces the height of the
potential barrier. Consequently, for a fixed angular momentum, a
spin $s$ particle in its lowest allowed angular momentum mode,
$l=s$, has a higher probability of overcoming the potential
barrier and reaching asymptotic infinity than particles with lower
spin and the same angular momentum. It should be noted, however,
that lower spin particles also possess modes with $l<s$, which can
alter this comparison when all allowed angular momentum modes are
taken into account. Thus, in the general case, the increase in
spin is accompanied by an increase in the minimum allowed angular
momentum $l$, so that, when all allowed modes are taken into
account, the transmission probability is expected to decrease with
increasing spin. This point can be illustrated by comparing the
effective potential for $s=l=2$ in Fig. \ref{fig:3a} with that for
$s=l=0$ in Fig. \ref{fig:2a}. Moreover, as the black hole loses
mass during the evaporation process and the quantum corrections
become increasingly significant, the height of the potential
barrier increases, thereby reducing the transmission probability
of the emitted particles, as shown in Fig. \ref{fig:3b}.

In summary, for a particle with a given spin, increasing the
angular momentum $l$ leads to a smaller contribution to the
corresponding Hawking emission spectrum, such that the mode with
the minimum allowed angular momentum, $l=s$, is expected to
dominate the spectrum of that particle species. Conversely, for a
fixed $l$, increasing the spin enhances the transmission
probability. However, in the general case, an increase in spin is
accompanied by an increase in the minimum allowed angular
momentum, $l=s$. Therefore, when all allowed $l$ modes are taken
into account, the contribution of a particle species to the
emission spectrum is expected to decrease with increasing spin. On
the other hand, the quantum corrections have a different effect on
the effective potential depending on the values of $s$ and $l$.
Except for the $s=l=0$ mode, for which they reduce the height of
the potential barrier, the quantum corrections increase the
effective potential for the other values of $s$ and $l$ considered
here. As the black hole approaches the final stages of
evaporation, these effects become increasingly pronounced: the
reduction of the potential for the $s=l=0$ mode becomes stronger,
while the increase in the potential for the other modes becomes
more significant. Consequently, one may expect the scalar
particles in the $s=l=0$ mode to make a relatively larger
contribution to the emission spectrum during the late stages of
evaporation compared with the early stages. These features are
examined in greater detail in the following sections.
\subsection{Bosonic greybody factors}\label{section 3b}
In the standard framework of black hole thermodynamics, a black
hole is not considered a perfect blackbody emitter, but rather a
greybody radiator. Although Hawking radiation exhibits a thermal
spectrum near the horizon, the emitted particles must propagate
through the effective potential barrier surrounding the black hole
before reaching an observer at infinity. This barrier, which
depends on both the properties of the emitted modes and the black
hole geometry, partially reflects the radiation back toward the
black hole and partially allows it to escape. The transmission
probability through this effective potential barrier is
characterized by the greybody factor. The study of greybody
factors is essential for determining the actual energy emission
rate and estimating the lifetime of evaporating black holes.
Moreover, since these factors depend on the spin of the emitted
particles as well as on the parameters and geometric properties of
the black hole, they can provide theoretical signatures for
distinguishing between different gravitational theories and
exploring possible effects of extra dimensions.

The greybody factor can be obtained by comparing the amplitude of
the incident wave before encountering the potential barrier with
the amplitude of the transmitted wave after crossing it. Near the
outer horizon and in the asymptotic region, the effective
potential vanishes, $V_{sl}(r)\rightarrow0$, and therefore the
radial equation \eqref{12} admits plane-wave solutions in these
two limits. The asymptotic behavior of the radial function can be
written as
\begin{equation}\label{18}
R_{sl}(r^*) = \left\{
\begin{array}{ll}
e^{-i\omega r^*}+\mathcal{R}e^{i\omega r^*}, & r^*\rightarrow-\infty\\
\mathcal{T}e^{-i\omega r^*}, & r^*\rightarrow+\infty
\end{array},
\right.
\end{equation}
where $\mathcal{R}$ and $\mathcal{T}$ denote the reflection and
transmission amplitudes, respectively. Conservation of the flux
implies that $|\mathcal{T}|^2 +|\mathcal{R}|^2=1$.

\begin{figure*}[t]
\centering
\begin{subfigure}{0.45\textwidth}
\centering
\includegraphics[width=\linewidth]{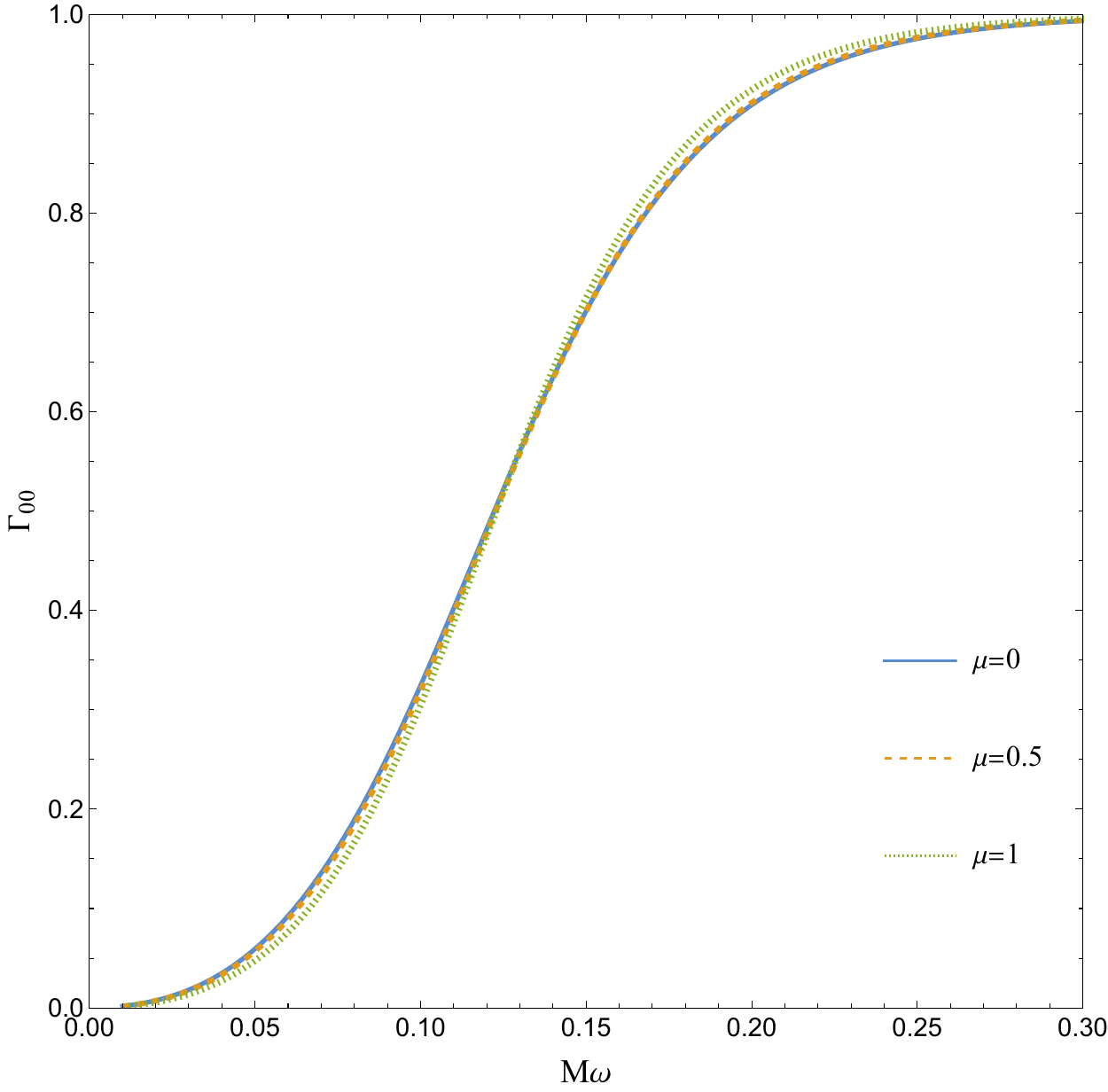}
\caption{$s=0,\ l=0$.}
\label{gb00}
\end{subfigure}
\hspace{0.05\textwidth}
\begin{subfigure}{0.45\textwidth}
\centering
\includegraphics[width=\linewidth]{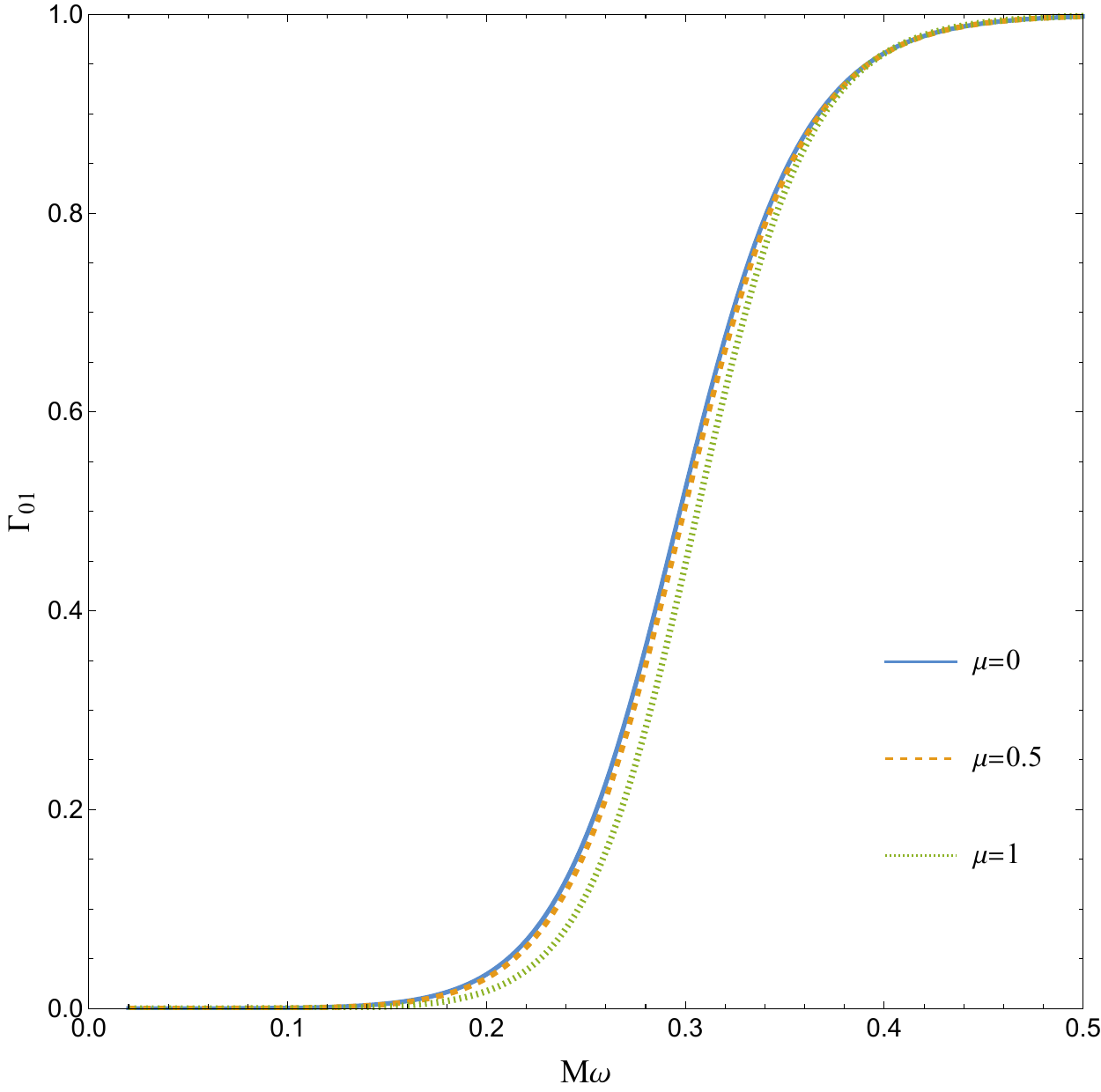}
\caption{$s=0,\ l=1$.}
\label{gb01}
\end{subfigure}

\vspace{0.3cm}

\begin{subfigure}{0.45\textwidth}
\centering
\includegraphics[width=\linewidth]{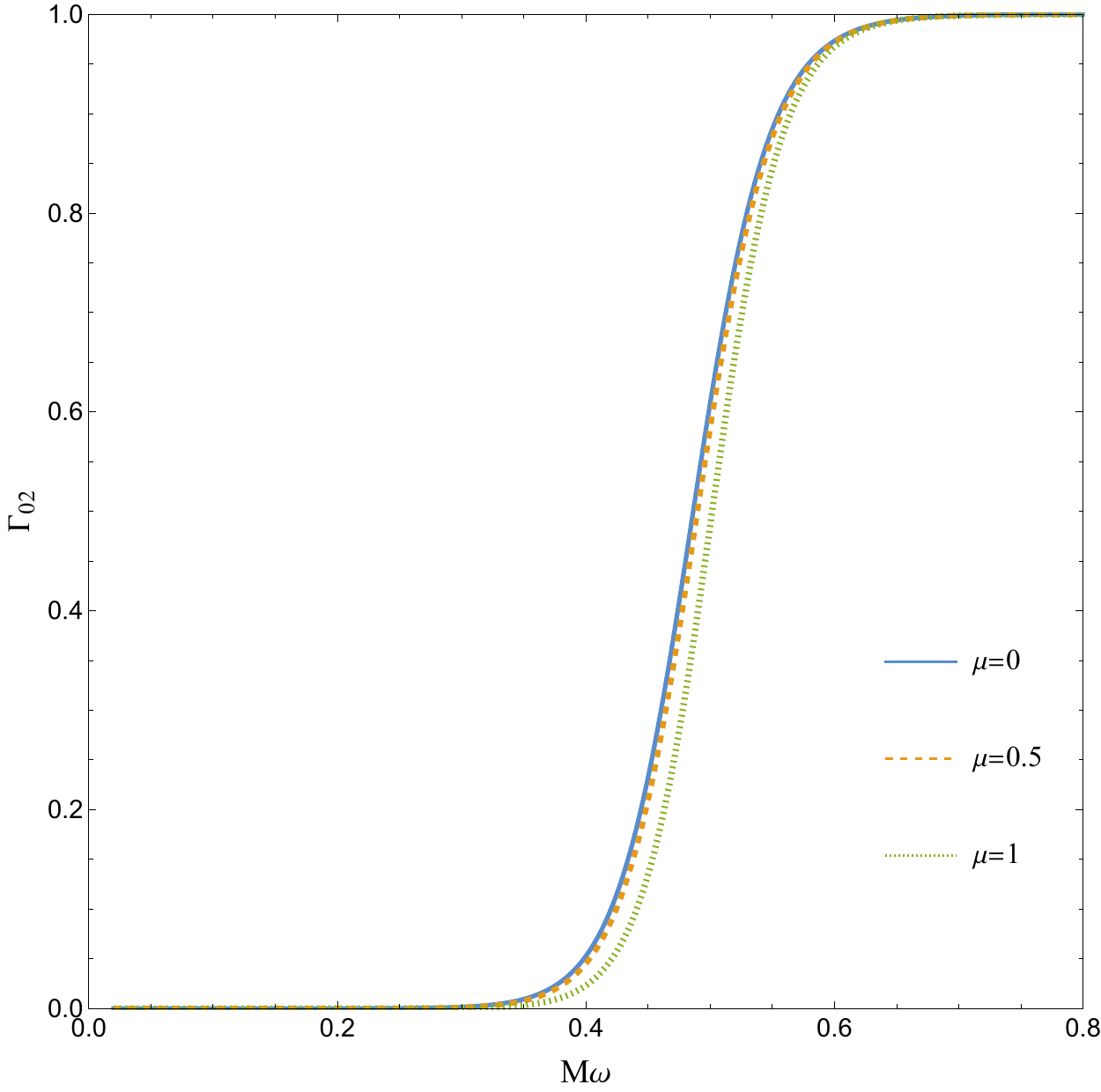}
\caption{$s=0,\ l=2$.}
\label{gb02}
\end{subfigure}

\caption{Greybody factors for massless scalar particles in the
quantum-corrected Schwarzschild spacetime, illustrating the
dependence on angular momentum modes $l$ and the quantum parameter
$\mu$.} \label{fig:4a}

\end{figure*}

\begin{figure*}[t]
\centering

\begin{subfigure}{0.45\textwidth}
\centering
\includegraphics[width=\linewidth]{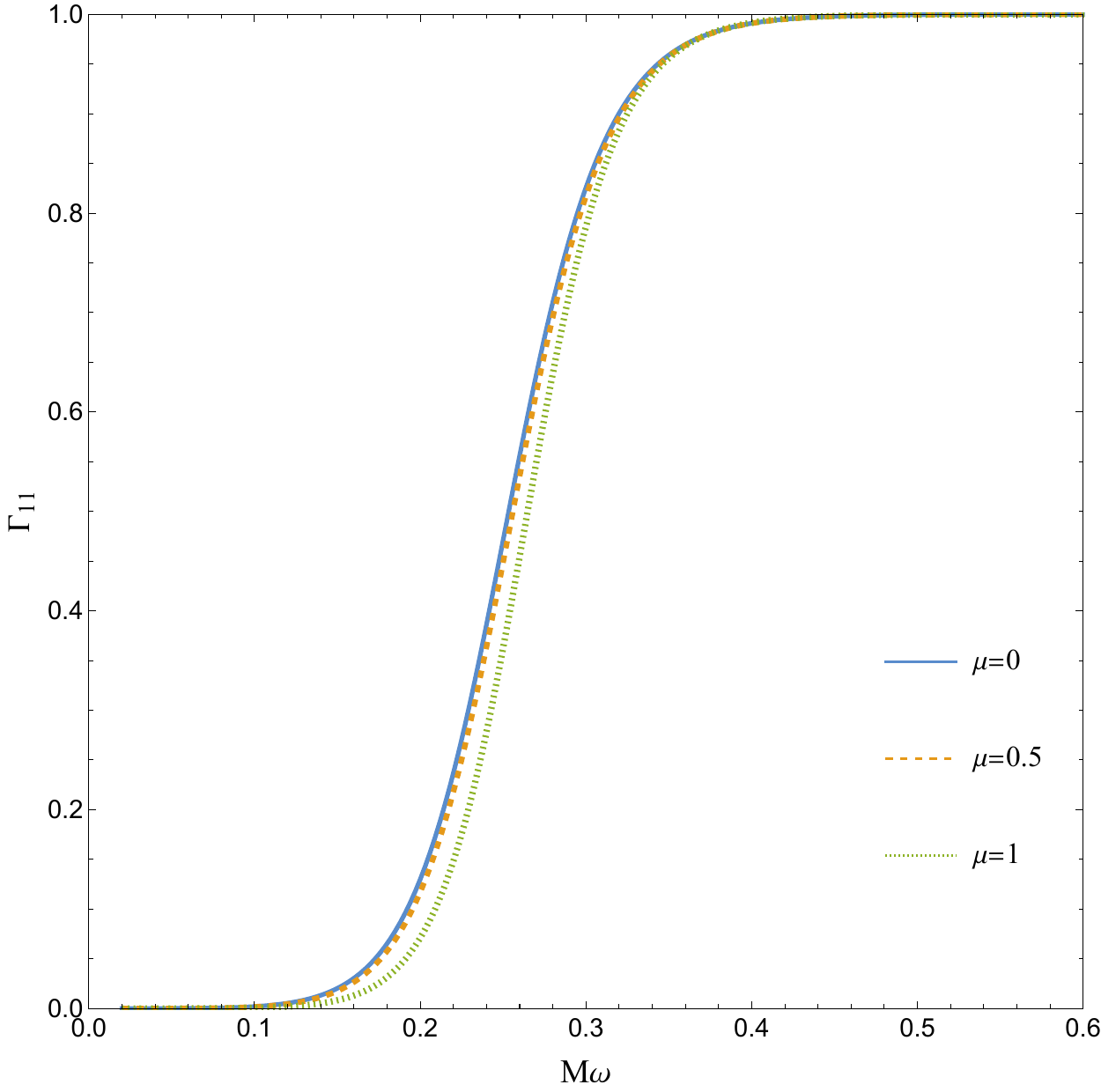}
\caption{$s=1,\ l=1$.}
\label{gb11}
\end{subfigure}
\hspace{0.05\textwidth}
\begin{subfigure}{0.45\textwidth}
\centering
\includegraphics[width=\linewidth]{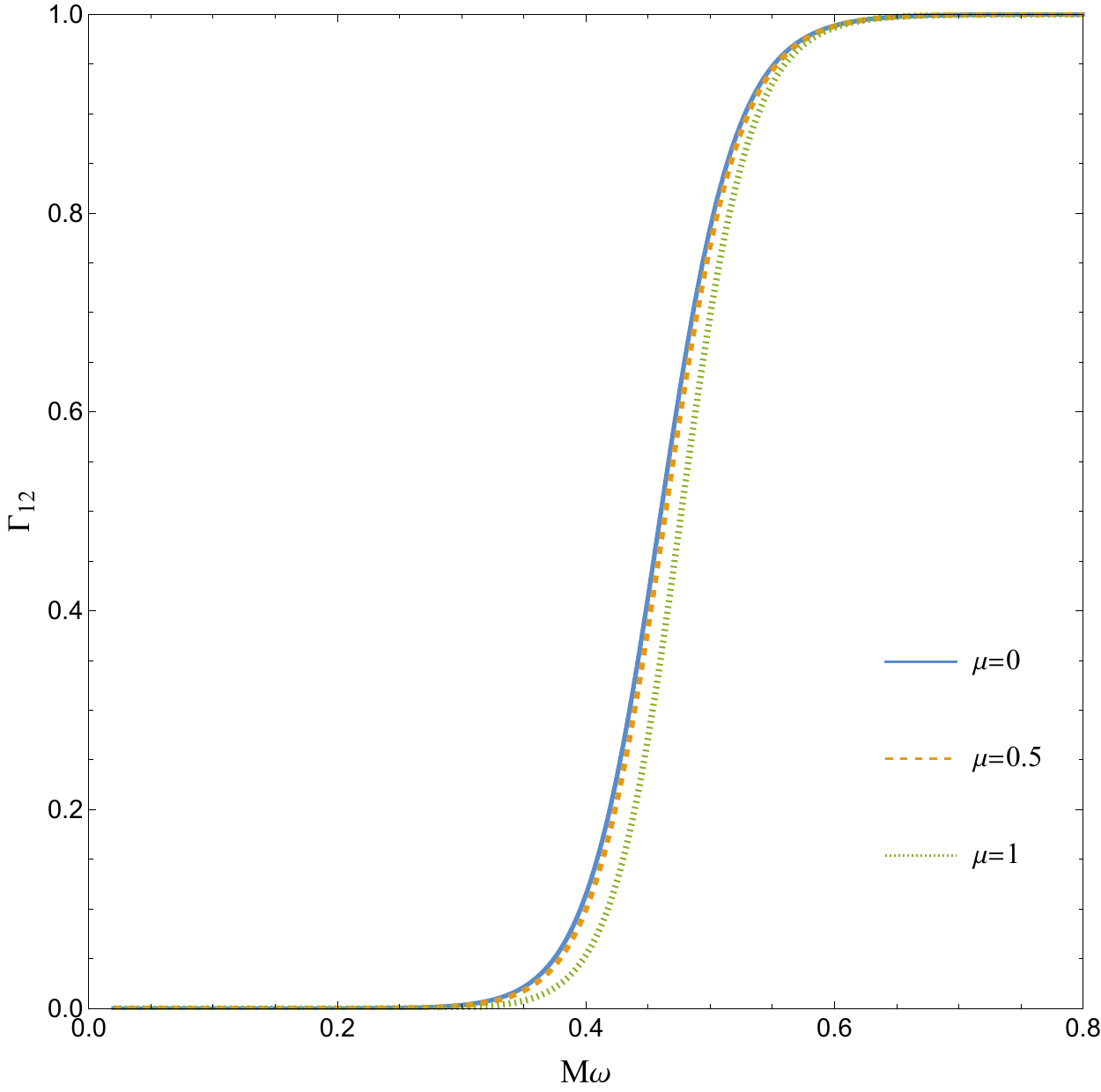}
\caption{$s=1,\ l=2$.}
\label{gb12}
\end{subfigure}

\vspace{0.3cm}

\begin{subfigure}{0.45\textwidth}
\centering
\includegraphics[width=\linewidth]{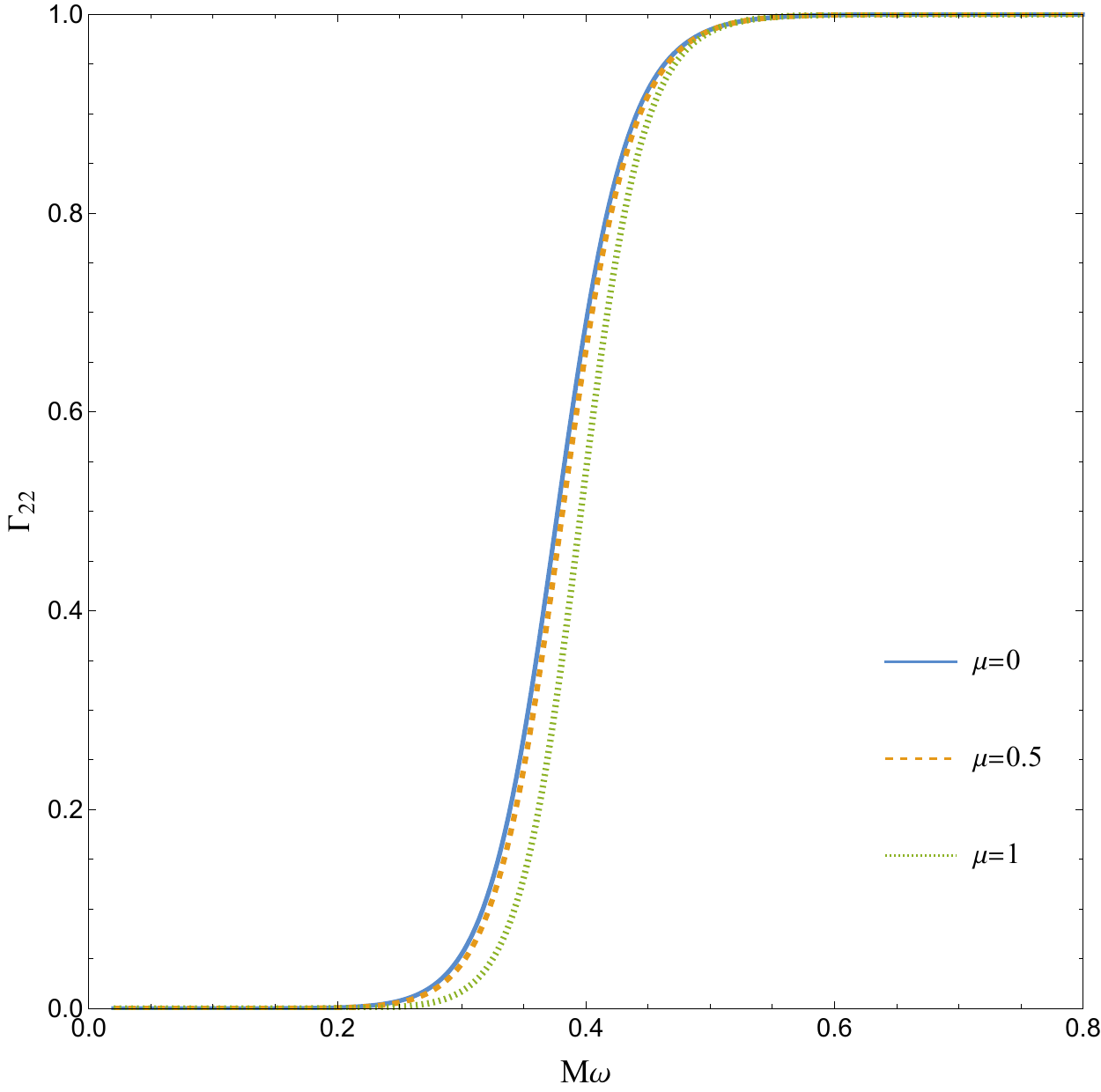}
\caption{$s=2,\ l=2$.}
\label{gb22}
\end{subfigure}

\caption{Greybody factors for photons ($l=1, 2$) and gravitons
($l=2$) in the quantum-corrected Schwarzschild spacetime, shown
for various values of the quantum parameter $\mu$.} \label{fig:4b}
\end{figure*}
Consequently, the greybody factor, which represents the
probability for an emitted particle to overcome the effective
potential barrier and reach spatial infinity, is given by
\begin{equation}\label{19}
\Gamma_{sl}(\omega)=|\mathcal{T}|^2.
\end{equation}
Therefore, the particle emission rate at spatial infinity for
massless bosonic particles with spin $s$ and angular momentum $l$
can be written as
\begin{equation}\label{20}
 \frac{d^2N_{sl}}{dt d\omega}=\frac{1}{2\pi}\frac{g_s(2l+1)\Gamma_{sl}(\omega)}{e^{\omega/T_H}-1},
\end{equation}
where $g_s$ represents the number of physical degrees of freedom
of the emitted particle. It is equal to $1$ for scalar particles
and $2$ for photons and gravitons, corresponding to their two
polarization states. The factor $2l+1$ accounts for the degeneracy
associated with the azimuthal quantum number $m$, which takes the
values $m=-l, ..., l$. The greybody factors are obtained by
numerically solving the radial Eq. \eqref{16} with the appropriate
boundary conditions at the outer horizon and spatial infinity. By
matching the asymptotic behavior of the radial solution in these
two regions, the transmission coefficient and consequently the
greybody factor can be extracted. Figure \ref{fig:4a} presents the
greybody factors for massless scalar particles for different values of the angular momentum $l$. For
scalar particles, as shown in Fig. \ref{fig:4a}, the greybody
factor decreases with increasing $l$ for a fixed particle energy
$\omega$. A similar behavior is observed for photons with $l=1$
and $l=2$, as shown in Figs. \ref{gb11} and \ref{gb12}. Therefore,
as expected from the stronger centrifugal barrier for larger
angular momentum modes, the transmission probability through the
effective potential barrier decreases as the angular momentum
increases. Equivalently, higher energies are required for these
modes to achieve comparable transmission probabilities. This
behavior also implies that, for a particle with spin $s$, the
dominant contribution to its emission spectrum is expected to come
from the lowest allowed angular momentum mode, namely $l=s$.

To investigate the role of spin in the greybody factor, we fix
$l=2$. As shown in Figs. \ref{gb02}, \ref{gb12}, and \ref{gb22},
for a given particle energy, the greybody factor increases with
increasing spin. For example, at $M\omega=0.4$, one obtains
$\Gamma_{02}\simeq0.05$, $\Gamma_{12}\simeq0.12$, and
$\Gamma_{22}\simeq0.69$. This behavior can be understood from Fig.
\ref{fig:3a} and Eq. \eqref{14}. The term proportional to
$(1-s^2)$ in the effective potential is positive for scalar
particles, vanishes for photons, and becomes negative for
gravitons. Consequently, increasing the spin reduces the effective
potential barrier and enhances the transmission probability.
Therefore, for a fixed angular momentum mode, higher spin
particles have a larger probability of transmission through the
effective potential barrier. However, when the full set of allowed
angular momentum modes is taken into account, the overall trend is
different. For lower spin particles, the existence of low angular
momentum modes leads to a smaller centrifugal contribution
$l(l+1)/r^2$, which facilitates the transmission of these modes
through the effective potential barrier. Since higher spin
particles do not possess such low $l$ modes (with the lowest
allowed value being $l_{min}=s$), their radiation spectrum is
generally suppressed compared with lower spin particles.

In addition to the spin and angular momentum of the emitted
particles, quantum corrections also modify the greybody factors,
as depicted in Figs. \ref{fig:4a} and \ref{fig:4b} . In general,
the greybody factor is also dependent on the magnitude of the
quantum corrections, characterized by the parameter $\mu$, such
that
\begin{equation}\label{21}
\Gamma_{sl}(\omega)\equiv\Gamma_{sl}(\omega; \mu).
\end{equation}
As the black hole evaporates and its mass decreases, the quantum
effects become increasingly significant (larger $\mu$).
Consequently, the transmission probability of low energy particles
through the effective potential barrier is suppressed. In other
words, compared with the early stages of evaporation, low
frequency modes have a smaller probability of escaping to infinity
during the final stages. Therefore, quantum corrections reduce the
greybody factors in the low energy regime.

As the particle energy increases, the difference between the greybody factors in the classical and quantum-corrected cases becomes smaller.
Beyond a certain frequency, i.e.,
\begin{equation}\label{22}
\Gamma_{\mu\ne0}(\omega_c)=\Gamma_{\mu=0}(\omega_c),
\end{equation}
quantum corrections instead enhance the greybody factors compared
with the classical case or the earlier stages of evaporation. As
shown in Figs. \ref{fig:4a} and \ref{fig:4b}, the crossing
frequency, at which the greybody factors for the quantum-corrected
cases ($\mu=0.5,1$) coincide with the classical value ($\mu=0$),
shifts toward higher frequencies as $l$ increases. This indicates
that the enhancement of the greybody factors due to quantum
corrections occurs at higher energies for larger angular momentum
modes, signifying the convergence toward the geometric optics
limit. Table \ref{tab1} presents the crossing frequencies for
$\mu=0.5$ and $\mu=1$, together with the corresponding greybody
factors for different modes. With increasing angular momentum $l$,
the crossing frequency moves to higher values, and the greybody
factor approaches unity, indicating the transition toward the
geometric optics regime where the potential barrier becomes
transparent for sufficiently energetic modes.

In summary, quantum corrections suppress the greybody factors in
the low energy regime while enhancing them in the high energy
regime.
\begin{table}[t]
\caption{Crossing frequencies and corresponding greybody factors
for different bosonic modes. The crossing frequency is defined as
the frequency at which the greybody factor in the
quantum-corrected case coincides with the classical value.}
\label{tab1}
\begin{ruledtabular}
\begin{tabular}{ccccc}
Mode $(s,l)$
& $\Omega_c(\mu=0.5)$
& $\Gamma_{sl}(\Omega_c)$
& $\Omega_c(\mu=1)$
& $\Gamma_{sl}(\Omega_c)
$
\\
\hline
$(0,0)$ & 0.15 & 0.70 & 0.13 & 0.55 \\
$(0,1)$ & 0.42 & 0.98 & 0.40 & 0.96 \\
$(0,2)$ & 0.70 & 0.99 & 0.68 & 0.99 \\
$(1,1)$ & 0.40 & 0.99 & 0.38 & 0.98\\
$(1,2)$ & 0.68 & 0.99 & 0.66 & 0.99 \\
$(2,2)$ & 0.55 & 0.99 & 0.53 & 0.99\\
\end{tabular}
\end{ruledtabular}
\end{table}
\section{Hawking radiation spectrum and total emitted power}\label{section 4}
Having determined the greybody factors, we can now investigate the
actual energy emission rate of the black hole for different
particle species. In this section, we analyze the effects of spin,
angular momentum, and quantum corrections on the Hawking radiation
spectrum and the total emitted power during the evaporation
process. In particular, quantum corrections become increasingly
important during the late stages of evaporation, where the
deviations from the classical Schwarzschild geometry become
significant. Within the effective semi-classical framework
considered here, these corrections provide the dominant
modifications to the radiation spectrum and evaporation dynamics.
\subsection{Energy emission spectrum}\label{section 4a}
The energy emission spectrum of a non-rotating and uncharged evaporating black hole for bosonic particles with spin $s$ and angular momentum $l$ is given by
\begin{equation}\label{23}
 \frac{d^2E_{sl}}{dt d\omega}=\frac{1}{2\pi}\frac{g_s(2l+1)\Gamma_{sl}(\omega)\omega}{e^{\omega/T_H}-1}.
\end{equation}
The total spectral power is then obtained by summing over all allowed spin and angular momentum modes, i.e,
\begin{equation}\label{24}
 \frac{d^2E}{dt d\omega}=\frac{1}{2\pi}\sum_{s, l}\frac{g_s(2l+1)\Gamma_{sl}(\omega)\omega}{e^{\omega/T_H}-1}.
\end{equation}

\begin{figure*}[t]
\centering
\begin{subfigure}{0.45\textwidth}
\centering
\includegraphics[width=\linewidth]{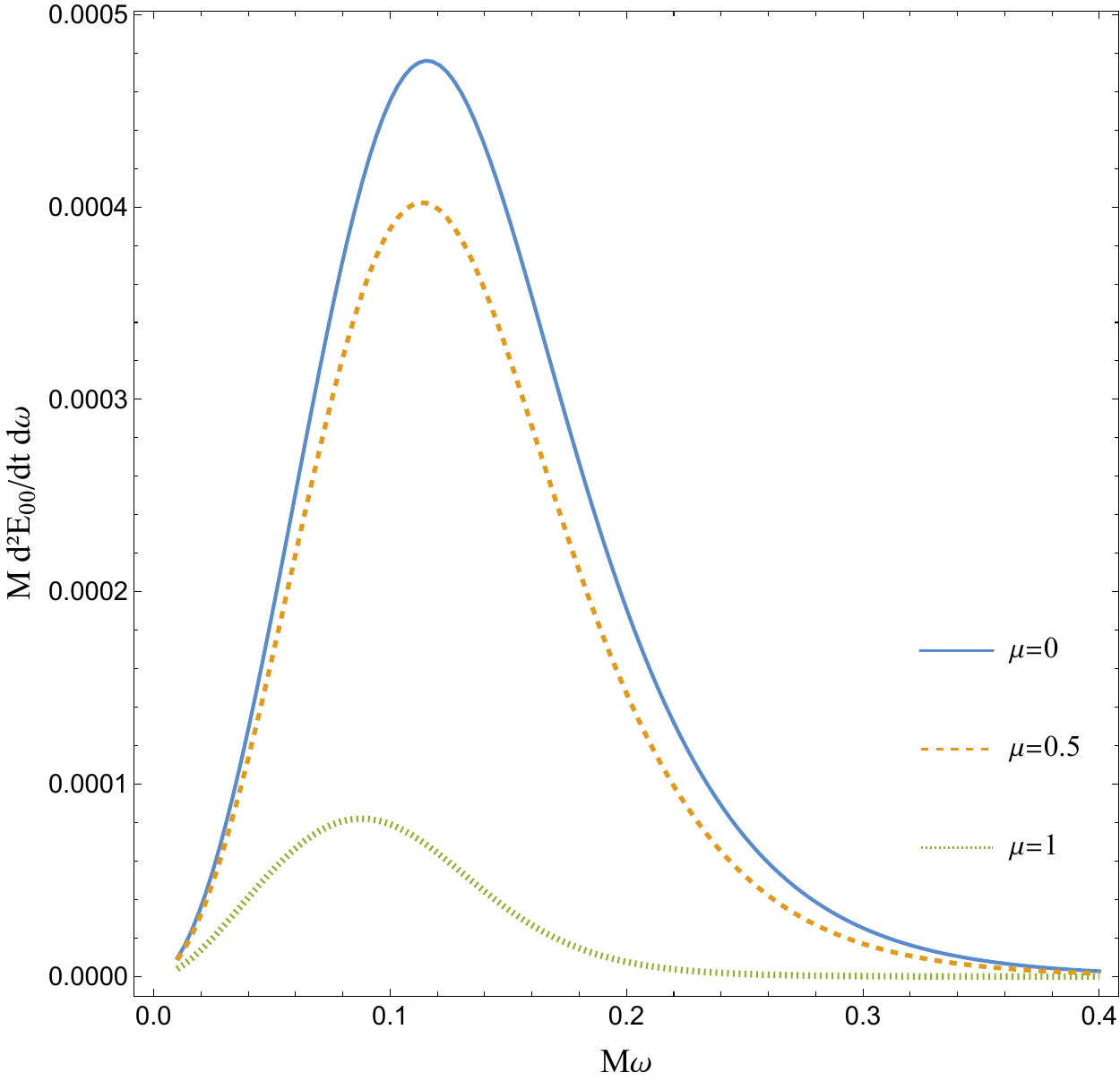}
\caption{$s=0,\ l=0$.}
\label{ef00}
\end{subfigure}
\hspace{0.05\textwidth}
\begin{subfigure}{0.45\textwidth}
\centering
\includegraphics[width=\linewidth]{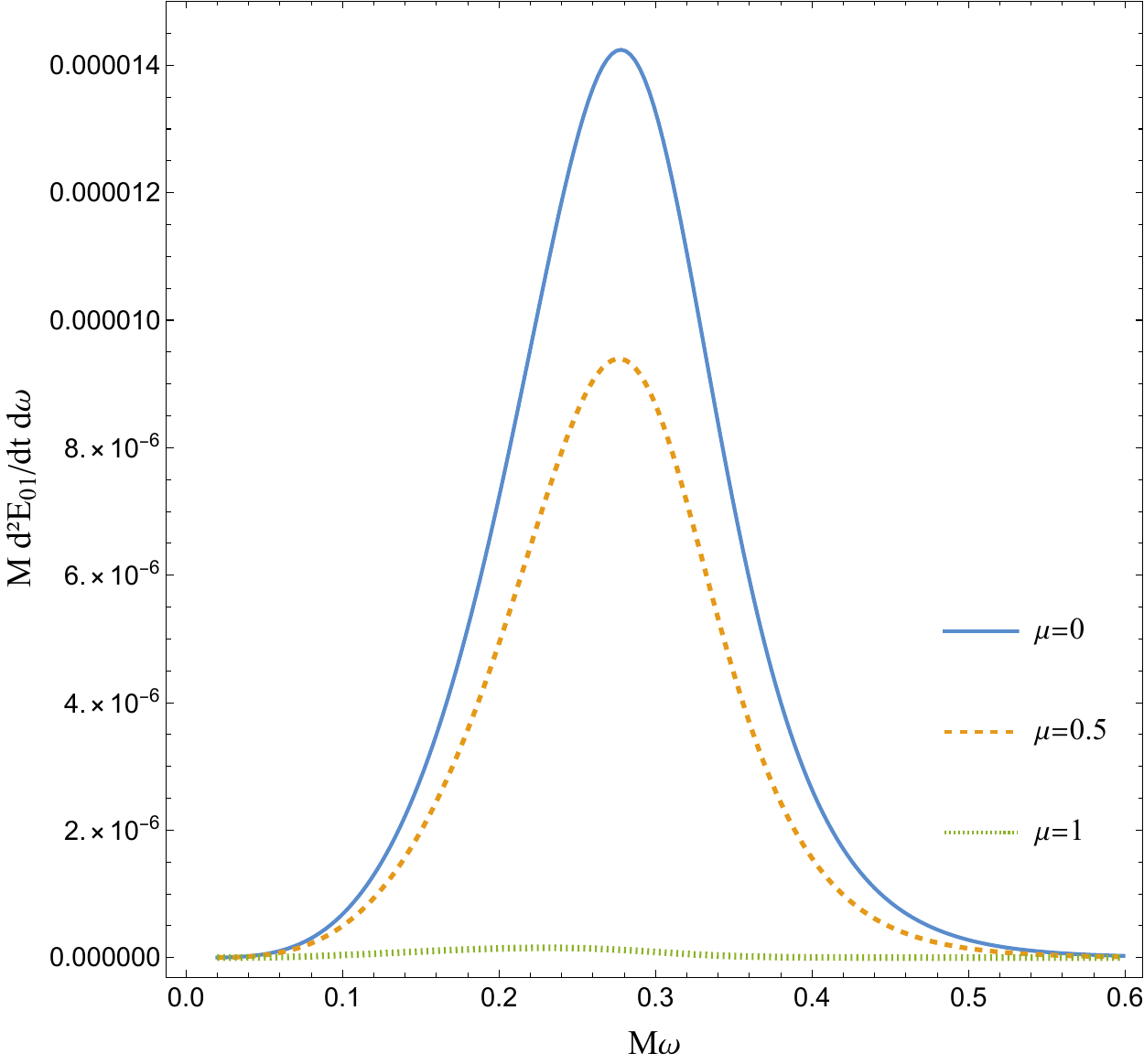}
\caption{$s=0,\ l=1$.}
\label{ef01}
\end{subfigure}

\vspace{0.3cm}

\begin{subfigure}{0.45\textwidth}
\centering
\includegraphics[width=\linewidth]{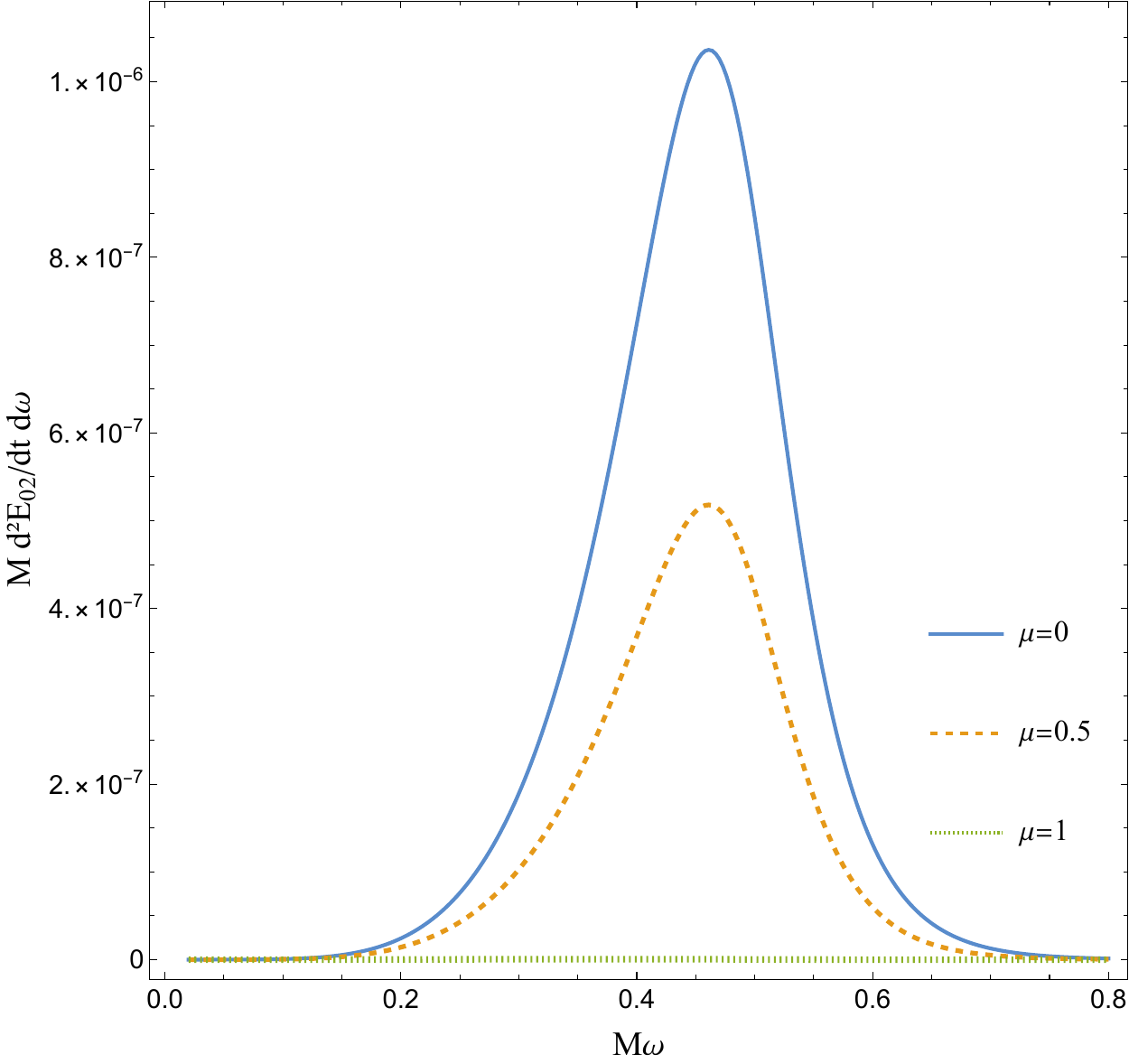}
\caption{$s=0,\ l=2$.}
\label{ef02}
\end{subfigure}
\caption{Energy flux of the massless scalar particles from a
quantum-corrected Schwarzschild black hole, plotted for various
angular momentum modes $l$ and different values of the quantum
parameter $\mu$.} \label{fig:5a}

\end{figure*}

\begin{figure*}[t]
\centering

\begin{subfigure}{0.45\textwidth}
\centering
\includegraphics[width=\linewidth]{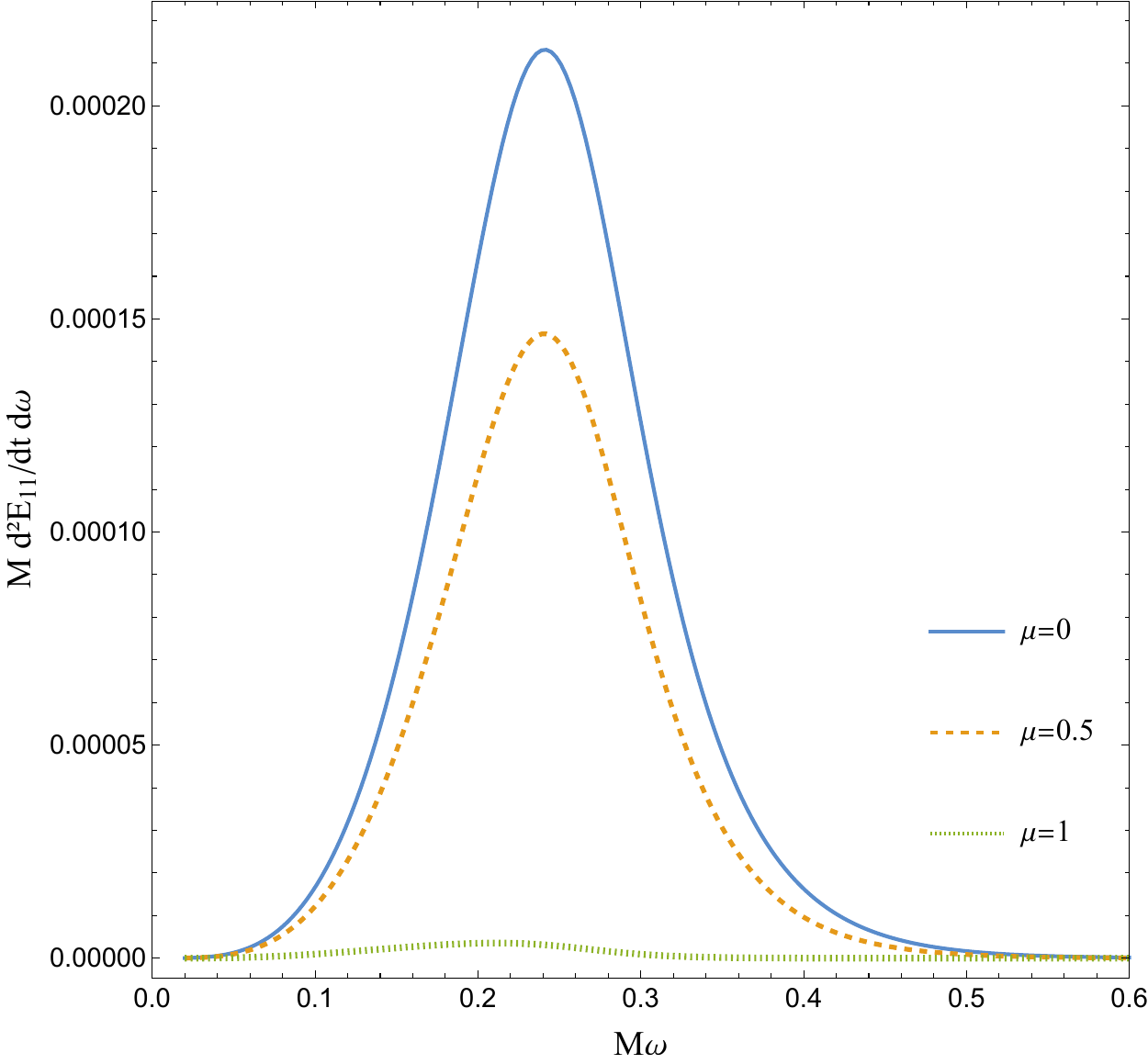}
\caption{$s=1,\ l=1$.}
\label{ef11}
\end{subfigure}
\hspace{0.05\textwidth}
\begin{subfigure}{0.45\textwidth}
\centering
\includegraphics[width=\linewidth]{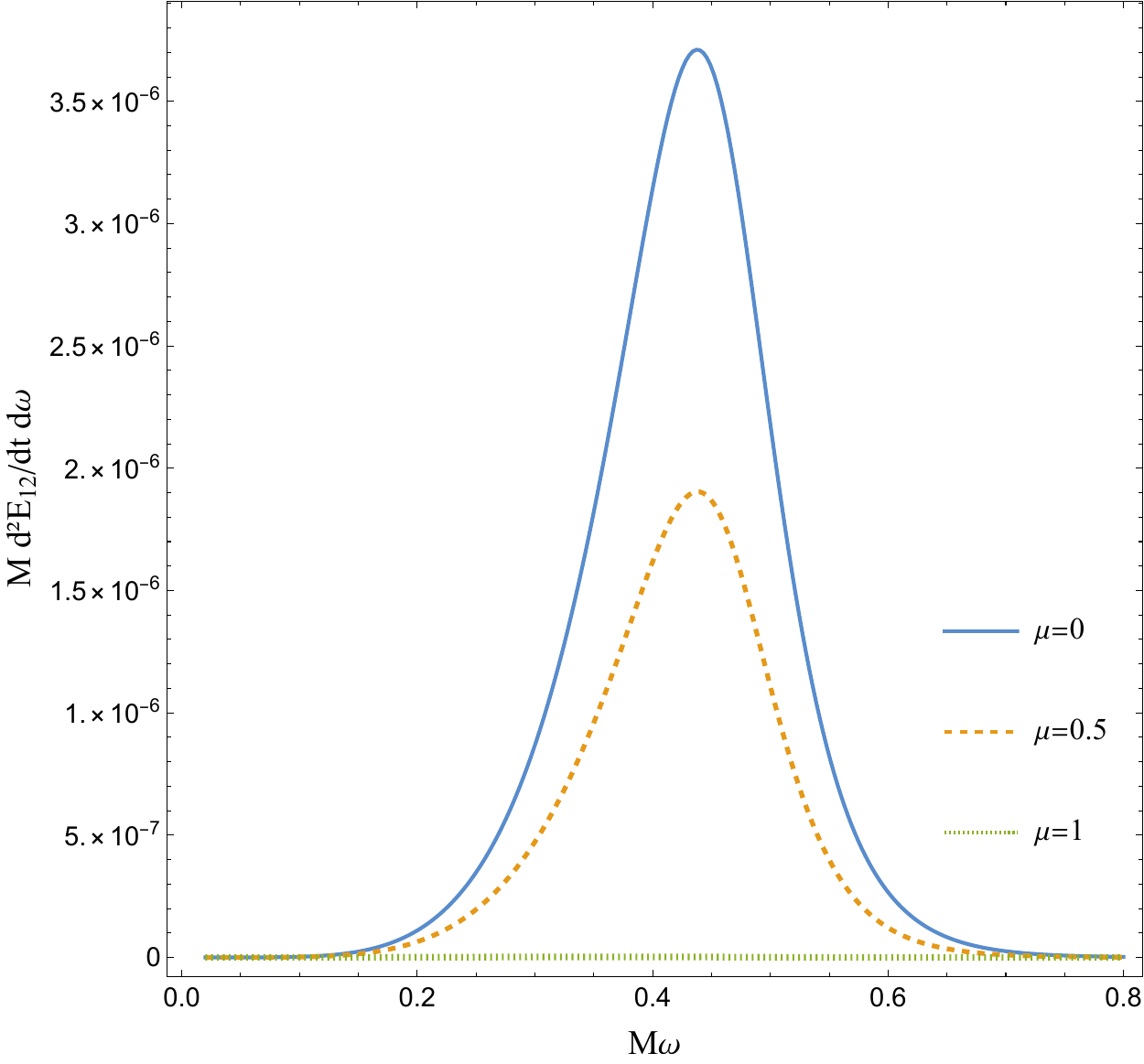}
\caption{$s=1,\ l=2$.}
\label{ef12}
\end{subfigure}

\vspace{0.3cm}

\begin{subfigure}{0.45\textwidth}
\centering
\includegraphics[width=\linewidth]{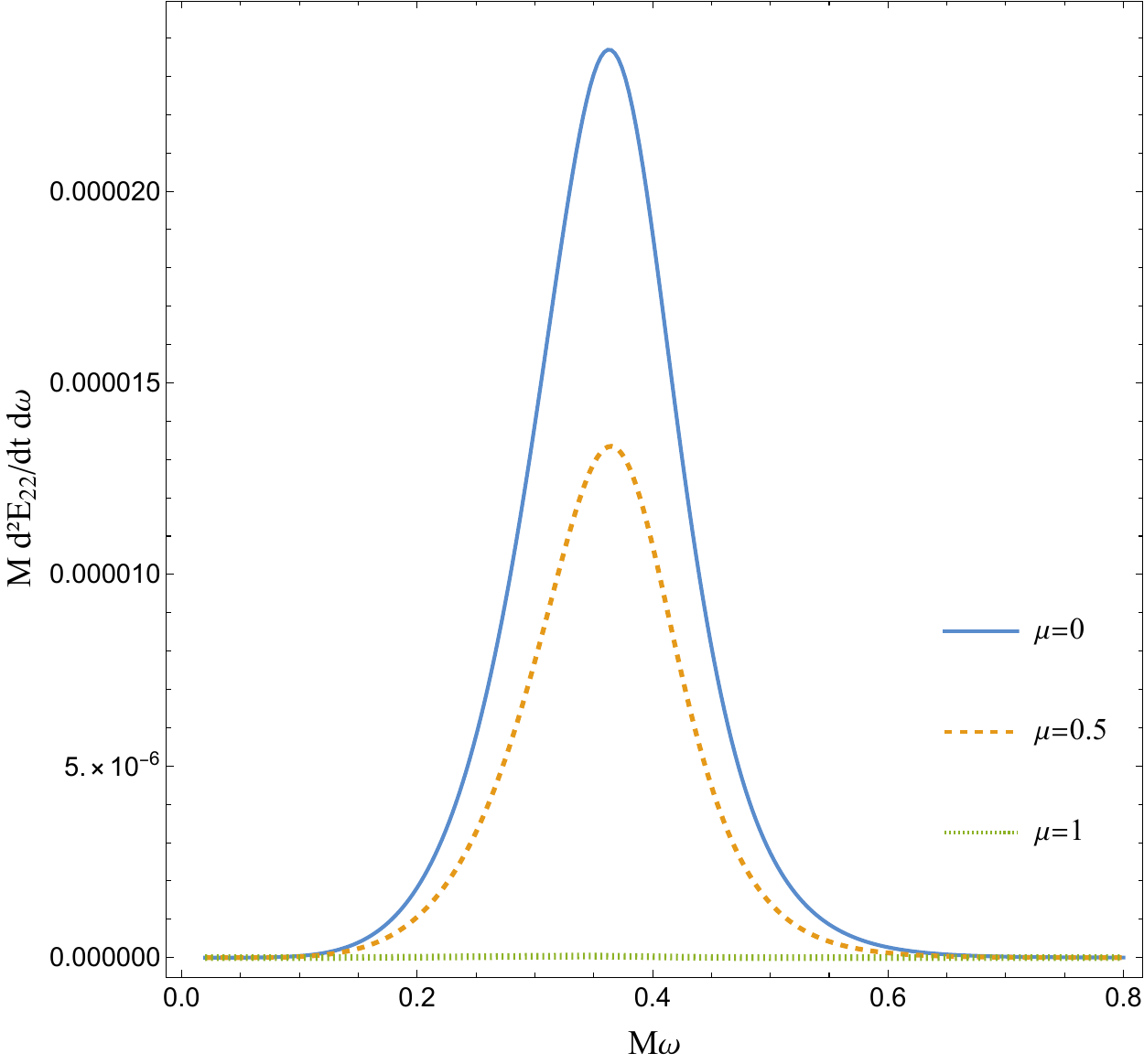}
\caption{$s=2,\ l=2$.}
\label{ef22}
\end{subfigure}
\caption{Energy flux of photons ($l=1, 2$) and gravitons ($l=2$)
from a quantum-corrected Schwarzschild black hole, plotted for
different values of the quantum parameter $\mu$.} \label{fig:5b}
\end{figure*}

The energy emission spectrum depends not only on the properties of
the emitted particles and their greybody factors, but also on the
Hawking temperature of the black hole. Since quantum corrections
modify the temperature behavior, particularly in the regime where
the effective quantum corrections become significant, they are
expected to leave imprints on the emitted radiation spectrum. The
energy spectra emitted by particles with different spins and
angular momenta are shown in Figs. \ref{fig:5a} and \ref{fig:5b}.
As can be seen in Figs. \ref{ef00}, \ref{ef01}, and \ref{ef02},
for scalar particles, increasing the angular momentum number $l$
suppresses the emitted energy. In particular, the height of the
spectral peak decreases, while its position shifts toward higher
energies. This behavior is a consequence of the increase in the
effective potential barrier with increasing angular momentum. The
same trend is also observed for photons with $l=1$ and $l=2$, as
shown in Figs. \ref{ef11} and \ref{ef12}.

For a fixed angular momentum $l=2$, Figs. \ref{ef02}, \ref{ef12},
and \ref{ef22} show that increasing the spin enhances the energy
flux and shifts the peak toward lower energies. Therefore,
gravitons with the minimum allowed angular momentum $l=2$,
corresponding to their dominant mode, contribute more
significantly to the black hole evaporation than scalar particles
and photons with the same angular momentum.

However, when the dominant modes of each particle species are
considered, the situation changes. Scalar particles and photons
have dominant contributions from $l=0$ and $l=1$ modes,
respectively, whereas gravitons start from $l=2$. Since the
centrifugal contribution $l(l+1)/r^2$ is larger for higher spin
particles due to their larger minimum allowed angular momentum,
the total contribution of higher spin particles to the Hawking
radiation spectrum is generally suppressed. Consequently, scalar
particles and photons are expected to contribute more
significantly than gravitons to the total emitted radiation.

Quantum corrections also affect the emitted energy flux through
their modifications of both the greybody factors and the Hawking
temperature. For large black holes, where quantum corrections are
negligible ($\mu=0$), the Hawking radiation spectrum is dominated
by effectively massless particles, including the three generations
of neutrinos and antineutrinos, photons, and gravitons \cite{gb1}.
In general, the emission of a particle with rest mass $m$ becomes
significant only when the black hole temperature satisfies
$T_H\gtrsim m$. For instance, a black hole with mass
$10^{20}\,\mathrm{g}$ has a Hawking temperature of
$105\,\mathrm{eV}$, which is below the rest mass of the lightest
charged Standard Model particles, namely electrons and positrons
($0.5\,\mathrm{MeV}$). Therefore, the radiation spectrum of such a
black hole contains only effectively massless particles.
\begin{table*}[t]
\centering \caption{The peak values of the energy emission rate
and the corresponding peak energies for different spin and angular
momentum modes at various values of $\mu$. The maximum values of
the spectral energy emission rate is given in units of
$\left(M_{\rm Pl}/M\right)E_{\rm Pl}$.} \label{tab2}
\begin{ruledtabular}
\begin{tabular}{ccccccc}
$(s,l)$
& \multicolumn{2}{c}{$\mu=0$}
& \multicolumn{2}{c}{$\mu=0.5$}
& \multicolumn{2}{c}{$\mu=1$}
\\[3pt]
& $M\omega_{\rm peak}$
& $\left(d^2E/dtd\omega\right)_{\rm max}$
& $M\omega_{\rm peak}$
& $\left(d^2E/dtd\omega\right)_{\rm max}$
& $M\omega_{\rm peak}$
& $\left(d^2E/dtd\omega\right)_{\rm max}$
\\[3.5pt]
\hline
$(0,0)$ & 0.12 & $4.7\times10^{-4}$ & 0.11 & $4\times10^{-4}$ & 0.09 & $8.2\times10^{-5}$ \\
$(0,1)$ & 0.28 & $1.4\times10^{-5}$ & 0.28 & $9.4\times10^{-6}$ & 0.23 & $1.5\times10^{-7}$ \\
$(0,2)$ & 0.46 & $10^{-6}$ & 0.46 & $5.2\times10^{-7}$ & 0.38 & $5\times10^{-10}$ \\
$(1,1)$ & 0.24 & $2.1\times10^{-4}$ & 0.24 & $1.5\times10^{-4}$ & 0.21 & $3.5\times10^{-6}$ \\
$(1,2)$ & 0.44 & $3.7\times10^{-6}$ & 0.44 & $1.9\times10^{-6}$ & 0.38 & $2.3\times10^{-9}$ \\
$(2,2)$ & 0.36 & $2.3\times10^{-5}$ & 0.36 & $1.3\times10^{-5}$ & 0.34 & $3.9\times10^{-8}$ \\
\end{tabular}
\end{ruledtabular}
\end{table*}
However, as the black hole loses its mass through evaporation and
its temperature increases, the emission of massive particles
becomes possible. For the case $\mu=0.5$, the black hole
temperature reaches a value of $2.3\times10^{14}\,\mathrm{TeV}$ ,
which is much larger than the rest masses of the Standard Model
particles. Therefore, the Hawking radiation spectrum can include
massive particle species. Since in this regime $T_H\gg m$, the
emission of particles with rest mass $m$ mainly occurs in the
ultrarelativistic regime, where their kinetic energy dominates
over their rest energy. Consequently, their rest masses can be
neglected with high accuracy. Therefore, for $\mu=0.5$, the zero
rest mass approximation remains valid for calculating the Hawking
radiation spectrum of massive particles.

It should be noted that, for $\mu=0.5$, the spectrum obtained from
the massless $s=1$ field equation should not be interpreted as
being restricted solely to photons. Since the Hawking temperature
is much larger than the rest masses of spin 1 particle species in
this regime, their emission occurs predominantly in the
ultrarelativistic limit. Therefore, the zero rest mass
approximation allows the $s=1$ spectrum calculated here to
approximately represent the contribution of all spin 1 Standard
Model particles. Also, due to the instability of some Standard
Model particles, a fraction of the initially emitted particles may
decay into other species, thereby modifying the observed Hawking
radiation spectrum. For example, the neutral pion, with a rest
mass of $135\,\mathrm{MeV}$, rapidly decays into two photons,
\begin{equation*}
\pi^0 \rightarrow \gamma+\gamma,
\end{equation*}
and consequently contributes to the observed photon spectrum
through secondary emission. Therefore, in the $\mu=0.5$ regime,
the photon spectrum does not correspond solely to directly emitted
photons, but can also contain photon contributions originating
from the decay of other particles. In addition, unstable particles
such as the Higgs boson, which is a spin 0 particle, can decay
into other Standard Model species and can contribute to the
secondary spectra of different particle species. However, such
secondary contributions from particle decays are neglected in the
present analysis.

In general, the spectrum of a given particle species can be
divided into two components: the primary spectrum, which
corresponds to the direct Hawking emission of that particle from
the black hole, and the secondary spectrum, which arises from the
decay of other particles initially produced by the black hole.
Furthermore, particles produced through Hawking radiation can
interact with each other, which may modify their contributions to
the radiation spectrum \cite{int1, int2}. For example, electrons
and positrons emitted by the black hole can undergo pair
annihilation and produce photons \cite{int3}. Consequently, when
such interactions are taken into account, the contribution of
electrons and positrons to the spectrum is reduced, while the
photon contribution is enhanced. Nevertheless, it has been shown
that interactions among the emitted particles have a negligible
impact on the Hawking radiation spectrum. Therefore, these effects
can be safely neglected in the present analysis.

In the final stages of evaporation, corresponding to $\mu=1$, the
Hawking temperature becomes very small according to Fig. \ref{HT}.
Therefore, similar to the early stages of evaporation ($\mu=0$),
the radiation spectrum is dominated by effectively massless
particles. As shown in Figs.  \ref{fig:5a} and \ref{fig:5b}, the
emission rate is strongly suppressed due to the significant
reduction of the black hole temperature. This suppression becomes
more pronounced for particles with larger spin and angular
momentum. In this regime, the scalar contribution with the
dominant $l=0$ mode becomes enhanced, while the contributions from
higher spin particles, including photons and gravitons, are
strongly suppressed. Therefore, as the black hole temperature
approaches zero and Hawking radiation gradually fades away, the
remaining weak radiation is expected to be dominated by massless
scalar particles with zero angular momentum.

In summary, after reaching the maximum temperature, quantum
corrections lead to a significant decrease in the Hawking
temperature, resulting in a strong suppression of the evaporation
rate. Consequently, the highly suppressed late stage Hawking
radiation is, to a very good approximation, dominated by the
massless scalar $s=l=0$ mode. In Table \ref{tab2}, the maximum
value of emitted energy flux and the corresponding energy at which
this maximum occurs are presented for particles with different
spins and angular momenta. The trends discussed above, including
the suppression of higher spin and higher angular momentum modes
and the enhancement of the scalar $s=l=0$ contribution in the late
stages of evaporation, are also evident in this table.
\subsection{Total emitted power and fields contributions}\label{section 4b}
By summing the spectral energy emission rate over all allowed
particle species and angular momentum modes, the total emitted
power (luminosity) of the black hole is obtained
\begin{equation}\label{25}
P=\int^\infty_{0}\frac{d^2E}{dt
d\omega}d\omega=\frac{1}{2\pi}\sum_{s,
l}g_s(2l+1)\int^\infty_{0}\frac{\Gamma_{sl}(\omega)\omega}{e^{\omega/T_H}-1}d\omega.
\end{equation}
Since the Hawking temperature is determined by the black hole
mass, the total emitted power is also a function of the black hole
mass. Therefore, the evaporation rate is expected to evolve
throughout different stages of the evaporation process. For
$\mu=0$, corresponding to the early stage of evaporation, the
total power carried by bosonic particles is found to be
\begin{equation}\label{26}
 P=4.03\times10^{55} \left(\frac{M_{Pl}}{M}\right)^2 \rm{erg\ s}^{-1}.
\end{equation}
The dominant contribution is 66.5\% from massless scalar
particles, followed by 30.1\% from photons and 3.4\% from
gravitons. For the case of $\mu=0.5$, the total power emitted in
the bosonic channels is equal to
\begin{equation}\label{27}
 P=3.06\times10^{54} \left(\frac{M_{Pl}}{M}\right)^2 \rm{erg\ s}^{-1},
\end{equation}
where 70.6\% of the total power is contributed by scalar
particles, 26.9\% corresponds to photons (more precisely, the spin
1 contribution, since the zero rest mass approximation is valid in
this regime), and 2.5\% is associated with gravitons. For $\mu=1$,
corresponding to the late stage of evaporation, the total power
contribution from massless bosonic species is given by
\begin{equation}\label{28}
 P=3.43\times10^{50} \left(\frac{M_{Pl}}{M}\right)^2 \rm{erg\ s}^{-1},
\end{equation}
where massless scalar particles, photons, and gravitons account
for 94.1\%, 5.8\%, and 0.1\% of the total emitted power,
respectively. In Table \ref{tab3}, the contributions of the spin
0, spin 1, and spin 2 fields to the bosonic Hawking radiation
spectrum are presented for the dominant angular momentum modes. As
can be seen, the overall contribution of higher spin fields to the
total emitted spectrum decreases. As the black hole becomes
smaller and the quantum corrections become stronger, the
contributions of spin 1 and spin 2 particles are further
suppressed, while the scalar contribution becomes dominant. This
behavior is consistent with Fig. \ref{fig:2}, where the effective
potential barrier for photons and gravitons increases due to their
nonzero angular momentum, whereas the scalar mode with $l=0$ does
not experience the centrifugal barrier and its effective potential
is reduced. Therefore, in the final stages of evaporation, the
dominant contribution to the mass loss of the black hole within
the considered bosonic sector is carried by scalar particles.
\begin{table}[t]
\centering \caption{Contribution (in \%) of different fields to
the total bosonic Hawking radiation power for the dominant modes:
scalar $(s=0,l=0,1,2)$, photon $(s=1,l=1,2)$, and graviton
$(s=2,l=2)$ at different values of $\mu$.} \label{tab3}
\begin{tabular}{c c c c}
\hline\hline
$\mu$ & Scalar (\%) & Photon (\%) & Graviton (\%)\\
\hline
0~({\rm classical}) & 66.5 & 30.1 & 3.4\\
0.5 & 70.6 & 26.9 & 2.5\\
1 & 94.1 & 5.8 & 0.1\\
\hline\hline
\end{tabular}
\end{table}
In Table \ref{tab4}, the contributions of the dominant modes to
the spectra of scalar and photon fields are presented. As can be
seen, increasing the angular momentum $l$ enhances the effective
potential barrier and consequently suppresses the contribution of
higher $l$ modes to the radiation spectrum of each field.
Furthermore, stronger quantum effects lead to a further
suppression of higher angular momentum modes. Therefore, in the
final stages of evaporation, the spectrum of each field is
dominated by the lowest allowed angular momentum mode, $l_{\rm
min}=s$.
\begin{table}[t]
\centering \caption{Relative contribution (in \%) of different
angular modes $l$ to the Hawking radiation power within each
field. The results are shown for scalar $(l=0,1,2)$ and photon
$(l=1,2)$ fields at different values of $\mu$.} \label{tab4}
\renewcommand{\arraystretch}{1.25}

\begin{tabular}{c c c c}
\hline\hline
Field & $\mu$ & $l$ & Contribution (\%)\\
\hline

\multirow{9}{*}{Scalar $(s=0)$}
& $0$~({\rm classical}) & 0 & 90\\
& & 1 & 9.75\\
& & 2 & 0.25\\
\cline{2-4}

& 0.5 & 0 & 91.9\\
& & 1 & 7.9\\
& & 2 & 0.2\\
\cline{2-4}

& 1 & 0 & 99\\
& & 1 & 0.99\\
& & 2 & 0.01\\

\hline

\multirow{6}{*}{Photon $(s=1)$}
& $0$~({\rm classical}) & 1 & 98\\
& & 2 & 2\\
\cline{2-4}

& 0.5 & 1 & 98.5\\
& & 2 & 1.5\\
\cline{2-4}

& 1 & 1 & 99.9\\
& & 2 & 0.1\\

\hline\hline
\end{tabular}
\end{table}

In summary, the quantum corrections significantly affect the
Hawking radiation properties by modifying both the greybody
factors and the black hole temperature. As the black hole
approaches the final stages of evaporation, the radiation power is
suppressed due to the decrease in temperature, while the
contribution of higher spin fields and higher angular momentum
modes becomes increasingly negligible. Consequently, the late
stage bosonic Hawking radiation is dominated by the scalar field
in the lowest angular momentum mode, indicating that the remaining
mass loss of the black hole is primarily carried by the $s=l=0$
mode.
\section{Evaporation process and black hole life-time}
The total emitted power determines the evaporation rate and the
mass loss rate of the black hole, which is described by
\begin{equation}\label{29}
\frac{dM}{dt}=-P(M).
\end{equation}
The rate and intensity of black hole evaporation depend on its
temperature and, consequently, on its mass. For instance, the
complete evaporation time of a solar-mass black hole is estimated
to be of the order of $10^{67}$ years. Therefore, the evaporation
rate of astrophysical black holes is extremely small and can be
neglected to a very good approximation. However, for primordial
black holes (PBHs), which are expected to have formed from density
fluctuations in the early Universe, evaporation can play a crucial
role. Since the evaporation rate increases significantly for
smaller black hole masses, the Hawking radiation process becomes
important for primordial black holes with sufficiently small
initial masses. It has been shown that the evaporation rate of
primordial black holes can be expressed as \cite{mac1, mac2}
\begin{equation}\label{30}
\frac{dM}{dt}=-5.34\times10^{25} F(M)M^{-2}\ \rm{g\ s}^{-1},
\end{equation}
where $F(M)$ depends on the types of particles emitted by the
black hole. As the black hole mass decreases, the Hawking
temperature increases, allowing the emission of particles with
larger rest masses. Therefore, the contribution of these massive
particle species should be taken into account in the evaporation
process.

By considering the contributions of all known Standard Model
particles, it has been shown that a primordial black hole with an
initial mass of $M=5\times10^{14}\,\mathrm{g}$ would be currently
undergoing the final stages of evaporation. Consequently,
primordial black holes with such initial masses have lifetimes
comparable to the present age of the Universe.

However, when quantum corrections are taken into account, a
different evaporation scenario emerges. Quantum effects cause a
significant reduction of the black hole temperature during the
final stages of evaporation and modify the behavior of the
function $F(M)$, leading to a strong suppression of the
evaporation rate. Using Eq. \eqref{29}, the black hole lifetime
can be calculated as
\begin{equation}\label{31}
\tau=\int_{M_{\rm min}}^{M_0}\frac{dM}{P(M)},
\end{equation}
where $M_0$ denotes the initial black hole mass and $M_{\rm min}$
represents the mass of the remnant produced at the end of
evaporation. It can be seen that in the limit $T_H\rightarrow0$
according to Eq. \eqref{25}, the emitted power approaches zero,
$P(M)\rightarrow0$, and consequently the lifetime diverges,
$\tau\rightarrow\infty$. By expanding $P(M)$ near the extremal
point, it can be shown that near the final evaporation stage,
\begin{equation}\label{32}
\tau\sim(M-M_{min})^{-3},
\end{equation}
indicating that the time required for the Hawking radiation to
become completely suppressed and for the remnant formation process
to be completed is extremely large. In other words, although a
quantum-corrected Schwarzschild black hole does not completely
evaporate, the timescale required to reach the final extremal
remnant state is significantly longer than the time required for
the complete evaporation of a classical Schwarzschild black hole.
This behavior is consistent with the third law of black hole
thermodynamics, which states that extremal black holes with
vanishing Hawking temperature cannot be reached through a finite
time physical process.
\section{Conclusions and discussions}\label{section 6}
Quantum corrections arising from vacuum polarization and conformal
anomaly effects modify the Schwarzschild geometry, leading to a
black hole structure characterized by two horizons: an outer
horizon representing a quantum-corrected extension of the
classical event horizon, and an inner horizon that emerges purely
due to quantum effects. This modified geometry is accompanied by
significant changes in the thermodynamic behavior of the black
hole.

Unlike the classical Schwarzschild case, where the Hawking
temperature increases continuously as the black hole loses mass,
the quantum-corrected black hole exhibits a bounded temperature
with a maximum value. Before reaching this maximum, the heat
capacity remains negative and the evaporation proceeds similarly
to the classical scenario. However, beyond the maximum
temperature, the heat capacity becomes positive, indicating a
transition toward a thermodynamically stable phase. During this
stage, further mass loss is accompanied by a decrease in
temperature rather than an increase. In the final stages of
evaporation, the inner and outer horizons approach each other and
merge into a degenerate horizon. Consequently, the Hawking
radiation becomes strongly suppressed, and the evaporation process
asymptotically approaches a zero temperature extremal remnant
within the effective semi-classical description. The thermodynamic
structure of the quantum-corrected black hole also leads to a
modification of the entropy beyond the classical
Bekenstein-Hawking area law. In particular, the entropy receives a
logarithmic correction induced by the quantum-corrected geometry.
Such logarithmic contributions appear in a variety of quantum
gravity approaches, including loop quantum gravity, Euclidean
quantum gravity, and other scenarios \cite{be1, be2, be3, be4,
be5, be6, be7}. In the present framework, these corrections arise
naturally from the modified thermodynamic relations associated
with the effective quantum-corrected metric, suggesting that
logarithmic modifications of the black hole entropy may represent
a generic signature of quantum gravitational effects.

In addition to modifying the thermodynamic properties, the
quantum-corrected geometry provides a nontrivial background for
studying the propagation of quantum fields and their emission
properties. In particular, the modifications of the spacetime
background directly affect the effective potential governing the
propagation of perturbations and consequently alter the Hawking
radiation spectrum. The effective potential experienced by the
emitted fields depends not only on their intrinsic properties,
such as spin and angular momentum, but also on the strength of the
quantum corrections. These corrections mainly modify the potential
profile in the near horizon region, where the quantum effects
become more pronounced. Although increasing the spin and angular
momentum generally increases the height of the effective potential
barrier, quantum corrections introduce a distinct behavior
depending on the angular momentum of the emitted modes. For modes
with nonzero angular momentum, quantum effects enhance the
potential barrier, whereas for scalar particles in the lowest
angular momentum mode ($l=0$), they reduce the barrier height. As
the black hole loses mass during evaporation and the quantum
corrections become increasingly important, these effects are
amplified. Therefore, quantum corrections affect the propagation
of different particle species in a non-universal manner, depending
on whether the emitted particles carry angular momentum or not.

The quantum corrections also leave distinct imprints on the
greybody factors and the transmission properties of the emitted
fields. As in the classical Schwarzschild case, increasing the
angular momentum barrier reduces the probability of transmission,
leading to smaller greybody factors for higher angular-momentum
modes. Moreover, increasing the spin of the emitted particles
further suppresses the greybody factors, since higher-spin fields
possess a larger minimum angular momentum and consequently
experience stronger effective potential barriers. The influence of
quantum corrections is found to be energy dependent: they suppress
the greybody factors in the low-frequency regime, while enhancing
them at higher frequencies. As the black hole mass decreases and
the quantum corrections become stronger, these deviations from the
classical behavior become increasingly pronounced. Interestingly,
the quantum-corrected and classical greybody factors intersect at
a characteristic crossing frequency. This frequency increases with
the angular momentum of the emitted modes and gradually approaches
the geometric optics limit in the high angular momentum regime.

The emitted energy flux spectrum also reflects the dependence of
Hawking radiation on the spin and angular momentum modes of the
radiated fields. In general, the contribution of each mode to the
radiation spectrum decreases as the spin and angular momentum
increase, reflecting the stronger suppression associated with
higher effective potential barriers. During the early stages of
evaporation, when the black hole mass is large and quantum
corrections are negligible, the bosonic radiation spectrum is
dominated by massless fields, including scalar particles, photons,
and gravitons. As the black hole approaches the regime of maximum
Hawking temperature, the temperature becomes sufficiently high
that all Standard Model particles can contribute to the radiation
spectrum. In this high temperature regime, massive particles are
emitted in an ultrarelativistic regime, allowing their rest masses
to be neglected and the zero rest mass approximation to remain
valid. Therefore, the spin 1 contribution in this regime should
not be interpreted as originating exclusively from photons, but
rather as the contribution of effectively massless spin 1 fields,
with other spin 1 bosonic degrees of freedom included in the
photon-like spectrum.

In the final stages of evaporation, as the Hawking temperature
decreases significantly due to quantum corrections, the radiation
spectrum becomes again dominated by truly massless particles,
similarly to the early evaporation phase. Moreover, the strong
suppression of the temperature near the extremal remnant leads to
a significant reduction of the emitted energy flux, while higher
spin and higher angular momentum modes become increasingly
suppressed. In contrast, the massless scalar field in the lowest
angular momentum mode provides the dominant contribution to the
late-time radiation spectrum. By summing over all spin and
angular-momentum modes, the total emitted power of the black hole
can be obtained. During the early stages of evaporation, where the
black hole mass is sufficiently large and quantum corrections are
negligible, the scalar field provides the dominant contribution to
the bosonic emission power, accounting for approximately
two-thirds of the total radiation. The remaining contribution
mainly originates from photons, while the contribution of
gravitons remains very small.

As the black hole temperature approaches its maximum value, the
scalar contribution becomes even more dominant, whereas the
contribution of spin 1 fields (including photons and other
effectively massless spin 1 bosonic degrees of freedom in the
Standard Model) and gravitons decreases. In the final stages of
evaporation, when quantum corrections strongly suppress the
Hawking temperature, the emitted radiation becomes significantly
weaker. In this regime, the photon contribution is substantially
reduced compared with the earlier stages of evaporation, the
graviton contribution becomes negligible, and the radiation
spectrum is almost completely dominated by massless scalar
particles in the lowest angular momentum mode ($l=0$). This
demonstrates that quantum corrections do not merely suppress the
overall evaporation rate, but also reshape the composition of the
Hawking radiation spectrum by progressively favoring the lowest
scalar mode near the extremal endpoint.

Finally, the evaporation timescale of the quantum-corrected black
hole was investigated by integrating the inverse emission rate
over the evolution of the black hole mass. Unlike the classical
Schwarzschild case, where the Hawking temperature continuously
increases as the black hole evaporates, the quantum-corrected
geometry leads to a vanishing temperature at the extremal
endpoint. As the black hole approaches this remnant state, the
emitted power is strongly suppressed and the evaporation rate
tends to zero, resulting in an extremely long timescale for the
completion of the evaporation process. In particular, near the
critical mass, the lifetime exhibits a divergent behavior,
indicating that the extremal remnant can only be approached
asymptotically. This behavior is consistent with the third law of
black hole thermodynamics, which states that an extremal
configuration with vanishing temperature cannot be reached through
a finite physical process. Therefore, within the effective
semi-classical framework considered here, the quantum-corrected
black hole does not completely evaporate within a finite time but
asymptotically evolves toward a stable extremal remnant. Similar
conclusions regarding the formation of an effective black hole
remnant through an infinite evaporation time have been obtained in
GUP-inspired models \cite{life1, life2}, where the evaporation
process also slows down as the black hole approaches its zero
temperature endpoint.

Several extensions of the present analysis can be considered in
future studies. A natural direction is to generalize the
quantum-corrected geometry to charged and rotating black holes,
where additional effects such as electromagnetic interactions,
frame dragging, and superradiant scattering may provide further
modifications to the greybody factors and evaporation process.
Another important extension is the investigation of fermionic
radiation, since the present work is restricted to massless
bosonic fields. Moreover, incorporating higher-order quantum
corrections into the effective geometry may reveal whether the
qualitative features found here, such as the existence of an
extremal remnant and the suppression of late-time evaporation,
remain robust beyond the leading-order approximation. Moreover,
primordial black holes have long been considered as possible
sources of high-energy astrophysical signals through their Hawking
radiation \cite{grb1, grb2, grb3}. In particular, the studies of
Hawking and Page \cite{grb1} showed that sufficiently light
primordial black holes could produce observable high-energy
emissions, including gamma rays, during the final stages of
evaporation. Since the quantum-corrected geometry studied here
significantly modifies the late-time evaporation dynamics and
suppresses the emission near the extremal endpoint, it would be
particularly interesting to examine how the absence of the
high-temperature final burst predicted in the classical
evaporation scenario affects the expected gamma-ray signatures of
primordial black holes and their observational constraints.

Furthermore, the extremal remnants emerging from the
quantum-corrected evaporation process may provide an interesting
connection to primordial black hole relic scenarios \cite{dm1,
dm2, dm3, dm4, dm5, dm6}. Investigating whether these stable
remnants can constitute a viable dark matter component requires a
detailed study of their cosmological production and abundance.
Exploring these cosmological and observational consequences,
including possible modifications of the expected gamma-ray
signatures and the viability of the resulting remnants as dark
matter candidates, represents an important direction for future
studies. Such investigations may help clarify the role of quantum
corrections in the final stages of black hole evaporation and
their possible connection to observable phenomena.
\acknowledgments{We are grateful to Shiraz University Research
Council.}

\bibliography{References}

\end{document}